\documentclass{article}
\usepackage[utf8]{inputenc}
\usepackage[margin=1in]{geometry}
\usepackage{authblk}

\title{Inference with generalizable classifier predictions}
\author[1]{Ciaran Evans\footnote{clevans@andrew.cmu.edu}}
\author[2]{Zara Y. Weinberg}
\author[3]{Manojkumar A. Puthenveedu}
\author[1]{Max G'Sell}
\affil[1]{Department of Statistics and Data Science, Carnegie Mellon University}
\affil[2]{Department of Biochemistry and Biophysics, University of California, San Francisco}
\affil[3]{Department of Pharmacology, University of Michigan}
\date{}

\usepackage{natbib}
\usepackage{graphicx}
\usepackage{amsmath}
\usepackage{amssymb}
\usepackage{amsthm}
\usepackage{mathtools}
\usepackage{multirow}
\usepackage{pdflscape}
\usepackage{bm}
\usepackage{bbm}
\usepackage{longtable}
\usepackage{booktabs}
\usepackage{float}
\usepackage[colorinlistoftodos]{todonotes}
\usepackage[font=small,labelfont=bf]{caption}

\usepackage[colorinlistoftodos]{todonotes}

\newcommand{\indep}{\rotatebox[origin=c]{90}{$\models$}}
\DeclareMathOperator*{\argmin}{arg\,min}

\usepackage[titlenumbered, ruled]{algorithm2e}
\SetKwProg{Init}{init}{}{}

\usepackage{hyperref}

\begin{document}
\maketitle

\begin{abstract} 
\noindent This paper addresses the problem of making statistical inference about a
population that can only be identified through classifier predictions.  The
problem is motivated by scientific studies in which human labels of a
population are replaced by a classifier.  For downstream analysis of the
population based on classifier predictions to be sound, the predictions must
generalize equally across experimental conditions.  In this paper, we formalize
the task of statistical inference using classifier predictions, and propose
bootstrap procedures to allow inference with a generalizable classifier.  We
demonstrate the performance of our methods through extensive simulations and a
case study with live cell imaging data.

\end{abstract}

\section{Introduction}
\label{sec:intro}

This paper studies statistical issues that arise when classifiers are used to
automate labor-intensive data collection in scientific pipelines.  With data
collection often requiring laborious human labels of objects or events, it has
become common to apply automated labeling techniques, in which a classifier is trained
on labeled examples to predict labels in new data \citep{norouzzadeh2018automatically, christiansen2018silico, caicedo2019nucleus}.
These classifier predictions, rather than ground truth labels, are then used
for statistical inference across groups and experimental conditions.

However, unless the classifier is perfect in all experimental settings, any
inference based on the classifier predictions must incorporate the additional
variability introduced by the classifier.  Furthermore, to make valid
comparisons across experimental units or conditions, the classifier must
exhibit the same performance across those units and conditions.  Since the
purpose of the study is often to show that two conditions are actually
different, this requirement is often unsatisified unless explicitly designed
for.

Motivated by experiments in which classifier predictions are the only feasible way to label large quantities of data, we present methodology for carrying out 
inference based on classifier-labeled data. We focus in particular on accounting for differences in data distribution between conditions.  
We outline considerations in designing a classifier beyond simple accuracy, and
define the necessary assumptions and models to perform valid statistical
inference. Our
work is inspired by the following case study from cellular biology.

\begin{figure}
\centering
\includegraphics[scale=0.3]{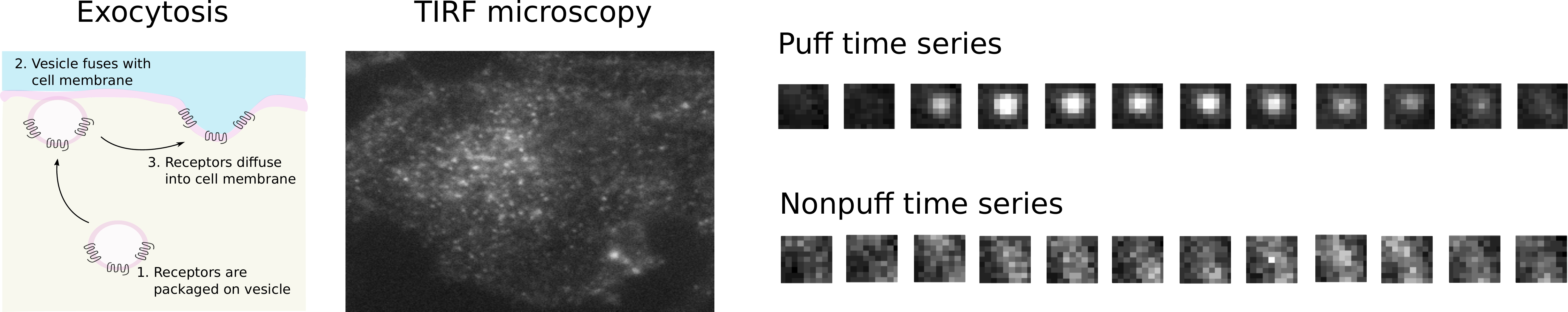}
\caption{Fluorescent proteins are used to study exocytosis with TIRF microscopy. \underline{Left}: Exocytosis regulated the concentration of receptor proteins on the cell surface by adding receptors through vesicle fusion. \underline{Center}: The surface of a cell in a TIRF microscopy image; bright spots correspond to concentrated clusters of receptors on the cell surface. \underline{Right}: Consecutive frames showing behavior of an exocytic event over time, in a 50Hz microscopy video. True exocytic events (``puffs'') have a characteristic pattern of diffusion over time. Other bright spots on the cell surface are not puffs, and do not show puff behavior.}
\label{fig:tirf}
\end{figure}

\textbf{Motivating example: Studying cellular transport through exocytosis.} 
Receptors on the cell surface play a crucial role
in a cell's response to external stimuli. These receptors---and thus the
corresponding responsiveness---are regulated in part by a process called
\textit{exocytosis}, which brings new receptors to the surface by packaging
them on bubbles of membrane, which then merge with the outer membrane of the cell
(Figure \ref{fig:tirf}) \citep{yu1993receptors, pippig1995receptors}.
Biologists can observe and measure exocytosis using 
\textit{total internal reflection fluorescence (TIRF) microscopy}, in which
individual exocytotic events manifest as bright ``puffs'' of fluorescence on
the cell surface, as
flourescence-tagged receptors are deposited and then diffuse (Figure \ref{fig:tirf})
\citep{axelrod1981tirf, sankaranarayanan2000gfp, rappoport2003tirf}. 
For simplicity, we'll refer to exocytic events as \emph{puffs} for this reason.
However, measuring these events is complicated by the fact that other objects
and processes in the cell can also manifest as bright spots in these TIRF
videos. Indeed, each cell in the data discussed in this paper typically has 5000 - 10000 detected events, and only about 5\% are puffs.  

\begin{figure}
\centering
\includegraphics[scale=0.4]{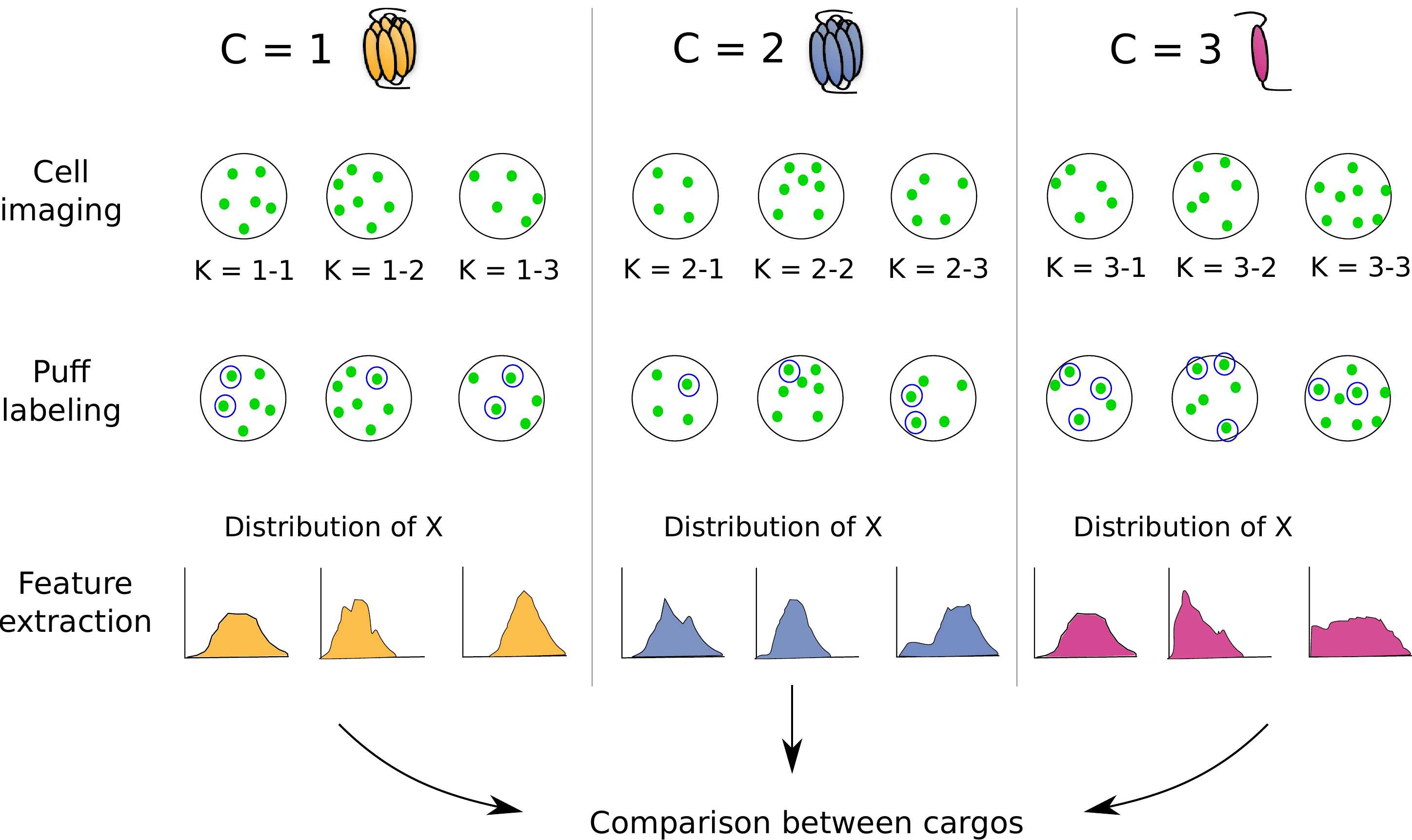}
\caption{Overview of an experiment to compare exocytic events between three different conditions ($C = 1, 2, 3$). For each condition, several cells $K$ are imaged with TIRF microscopy, and bright spots are identified. These detected events are then labeled as puffs or nonpuffs, and a feature $X$ is recorded for each puff. The distribution of $X$ is then compared across conditions.}
\label{fig:workflow}
\end{figure}

A typical experiment compares puff behavior for several different surface receptors or experimental conditions, resulting in a hierarchical structure common to biological experiments, shown in Figure \ref{fig:workflow}.  When TIRF microscopy images are labeled by hand, statistical inference between conditions is straightforward:
\begin{enumerate}
\item TIRF microscopy is performed for each cell
\item Researchers use characteristic patterns of puff appearance to identify puffs \citep{logan2017exocytosis, kou2019exocytosis}
\item Features describing puff behavior (e.g., how long
diffusion takes, what diffusion looks like, etc.) are recorded \citep{yudowski2006distinct, bowman2015cell, bohannon2017exocytosis}
\item These features are compared across conditions
\end{enumerate}

However, the volume of detected events in each cell makes manual labeling challenging. Automated labeling is therefore valuable, using a
classifier trained to predict whether a bright spot is a puff or nonpuff. These predictions are then used for inference in place of hand labels.

\textbf{Problem statement.} Inspired by this motivating example, we consider
data with the following structure, 
notated $\{(V_i, Y_i, C_i, K_i)\}_{i=1}^n$, where $Y \in
\mathcal{Y}$ is the \textit{unobserved} true label, $V$ is a set of observed
covariates, $C \in \mathcal{C}$ is experimental condition, and $K \in \mathcal{K}$ is a grouping variable nested within $C$ that captures the hierarchical structure of the data. For instance, in the TIRF microscopy example, $Y \in \{0, 1\}$ denotes
whether each event is a puff, $V$ is a set of features derived from the
microscopy images (either specially designed, or created by methods like
convolutional neural networks), $C$ denotes condition/receptor type, and $K$ denotes the cell
(Figure \ref{fig:workflow}). More generally, $C$ may indicate an experimental
condition, and $K$ a repetition of that experiment. We present the problem in
this hierarchical setting because it is most common, and note that our methods can
be simplified when the hierarchical structure does not apply.

Comparing experimental conditions $C \in \mathcal{C}$ often involves asking one or both of the following questions:
\begin{enumerate}
\item How does label prevalence $P(Y = y | C = c)$ vary for conditions $c \in \mathcal{C}$?

\item How does the conditional distribution of $X|Y = y, C=c$ vary for conditions $c \in \mathcal{C}$, where $X \in \mathbb{R}$, $X \subset V$, denotes a feature of interest?
\end{enumerate}
\textit{Crucially, $Y$ is unobserved in new experimental data.} To perform inference with unlabeled data, we have access to \textit{labeled} training data $\{(V_j', Y_j', C_j', K_j')\}_{j=1}^m$, with the same set of labels $Y_j' \in \mathcal{Y}$, but different conditions $C_j' \in \mathcal{C}'$ and $K_j' \in \mathcal{K}'$. 
As new studies and experiments typically investigate different conditions, training and test data necessarily come from different experimental conditions and groups ($\mathcal{C}' \cap \mathcal{C} = \emptyset$ and $\mathcal{K}' \cap \mathcal{K} = \emptyset$). Therefore, when making predictions with the training data, we must consider possible differences in the data distribution between training $\{(V_j', Y_j', C_j', K_j')\}_{j=1}^m$ and test $\{(V_i, Y_i, C_i, K_i)\}_{i=1}^n$.

\textbf{Contributions.} This paper makes the following contributions:
\begin{enumerate}

  \item We explicitly define the statistical task of downstream inference
    about label prevalence $P(Y=1|C=c)$ and the
class-conditional means $\mathbb{E}[X|Y=y, C=c]$ using
    classification predictions.  We codify the different sets of assumptions required to enable
    meaningful inference in this setting.
  \item We develop semiparametric bootstrap methods for making downstream inference with classifier predictions, which
    properly incorporate variance when the required generalizability assumptions hold.
  \item We demonstrate the use and performance of these methods in both
    simulation and in a
detailed case study with TIRF microscopy data, where we describe the process of
constructing a generalizable classifier, checking assumptions for valid
statistical inference, and creating bootstrap confidence intervals.
 \item Through our case study, we provide practical advice on feature and classifier construction in
    order to satisfy the assumptions necessary for downstream inference.
\end{enumerate}

In Section \ref{sec:background}, we describe previous literature on automatic
labeling and generalizable classifiers. In Section \ref{sec:inference}, we
present our bootstrap methods for inference with unlabeled data, and the
different assumptions required. We demonstrate the performance of our methods
on simulated data in Section \ref{sec:simulations}, and discuss possible
modifications of our algorithms for different scenarios. Finally, in Section
\ref{sec:case-study}, we perform inference in a case study with real TIRF
microscopy data.

\section{Background}
\label{sec:background}

\subsection{Automated labeling in scientific studies}

Automated labeling is useful in a wide range of applications, often with large
image datasets. For example, \cite{norouzzadeh2018automatically}  developed a
classifier to identify animal species in camera trap images, as well as the
number of animals in each picture and a description of their actions. The
classifier was trained on millions of images from the Snapshot Serengeti
dataset \citep{swanson2015snapshot}, using convolutional neural networks. In
ecology, large-scale classifiers have also been used to label deforestation
\citep{maretto2020spatio} and pest infestations \citep{rammer2019harnessing}.
Automated labeling is also common in cell biology, where microscopy can produce
thousands of images a day, which need to be annotated to identify nuclei
\citep{caicedo2019nucleus}, cell state and type \citep{christiansen2018silico}, cell health phenotypes \citep{way2021predicting},
and protein localization \citep{kraus2017automated}.

In each of these examples, it is important that classifiers generalize, or
\textit{transfer}, to new data. If predictions are not robust to changes in the
distribution of input data, classifiers can fail when applied to new settings
\citep{pan2009survey, quinonero2009dataset}. It has therefore become common for
researchers constructing automated labeling systems to design classifiers which
they expect to transfer. For example, \cite{norouzzadeh2018automatically}
describe how their method can be updated for new camera trap locations with
only a small amount of additional data.

\subsection{Generalizable classifiers}
\label{sec:generalizable-background}

A typical assumption in supervised learning is that training and test data come
from the same distribution, which allows classifier predictions to be
meaningful on new data. In practice, however, training and test data often come
from different distributions, and so assumptions on the nature of distributional change
are needed to understand how classifiers to generalize to new data.

Typically, we say that a classifier trained on data from condition $C = c'$, using the covariates $V$, generalizes to a new condition $C = c$ if $P(Y = y | V = v, C = c) = P(Y = y | V = v, C = c')$ \citep{subbaswamy2019preventing}. In this case, the features $V$ satisfy the \textit{covariate shift} assumption \citep{bickel2009discriminative, gretton2009covariate}: the marginal distribution of $V$ may change, but the conditional distribution of $Y|V$ remains the same. Under covariate shift, predictions can be applied directly to new data, or the classifier can be re-trained on weighted training data to be more efficient at risk minimization on test data \citep{shimodaira2000improving, sugiyama2008direct}. However, not all features typically satisfy the covariate shift assumption, as we expect systematic differences between conditions to appear in some features. Therefore, we write $V = \{X, Z, U\}$, where $X \in \mathbb{R}$ is a feature of interest for inference, $U$ is a set of unused features, and $Z \in \mathbb{R}^d$ is a subset of covariates which satisfy the covariate shift assumption: 
\begin{align}
P(Y = y | Z = z, C = c) = P(Y = y | Z = z, C = c'), \ \text{ for all } c, c'.
\end{align}
Ideally, $Z$ is also \textit{sufficient} to capture the label information, so that $P(Y = y | X, U, Z, C) = P(Y = y | Z)$ \citep{peters2016causal, kuang2018stable}. 

The strategy of identifying features that allow generalizability is common, and there are a variety of techniques. For example, \cite{magliacane2017domain} and \cite{rojas2018invariant} use variable selection techniques to identify a subset of predictors for which covariate shift holds, and \cite{peters2016causal} performs hypothesis tests on the relationship between the predictors and the response. These methods are inspired by causal inference and causal discovery, as are \cite{subbaswamy2019preventing}, who represent the data generating process explicitly with a causal graph and use the graph to identify stable predictors. \cite{kuang2018stable} also use ideas from causal inference, in particular balancing models which use weights to account for differences in covariate distributions across environments, to identify a subset of covariates which generalize. More broadly, other authors have developed regularized regression methods to learn features which are common to multiple environments \citep{argyriou2007spectral, argyriou2008convex}.

In some cases, however, it may not be possible to identify an appropriate subset of features $Z$ for which covariate shift holds \citep{subbaswamy2019preventing}. For example, if $Z|Y = y, C= c \ \overset{d}{=} \ Z|Y = y, C = c'$, but the prevalence $P(Y = y | C = c) \neq P(Y = y | C=c')$, then predicted probabilities will not be calibrated for all $c \in \mathcal{C}$. This is the \textit{label shift} scenario, in which the marginal distribution of labels changes, but the conditional distributions of features remain the same. Fortunately, there are a variety of methods for detecting and correcting for label shift, which allow predictions to be easily adjusted in this setting \citep{saerens2002adjusting, storkey2009training, lipton2018detecting, garg2020unified}.

To accomodate situations like label shift, in this manuscript we consider a classifier trained on features $Z$ to \textit{generalize} if label probabilities $P(Y=y | Z=z, C=c)$ for each condition $c \in \mathcal{C}$ can be estimated from labeled training data $\{(Z_j', Y_j', C_j', K_j')\}_{j=1}^m$ and unlabeled test data $\{(Z_i, C_i, K_i)\}_{i=1}^n$---either by directly applying a classifier, in the case of covariate shift, or by correcting classifier predictions, like in the case of label shift. The existence of appropriate features $Z$ is assumed; in practice, there are a variety of methods for identifying $Z$, as discussed above, and in our case study on TIRF microscopy data we use simple exploratory techniques.

\section{Inference with generalizable predictions}
\label{sec:inference}

We are interested in two inference questions using the classifier labels.
First, how label prevalence, $P(Y=y|C=c)$, differs across conditions $C$.  Second,
how the conditional distribution of a feature of interest $X$, $X|Y=y, C=c$,
differs across conditions $C$. As the labels $Y$ are unobserved for the new data, we use a classifier
$\mathcal{A}$ trained on labeled training data $\{(Z_j', Y_j', C_j',
K_j')\}_{j=1}^m$. For simplicity, we'll assume that $Y \in \{0, 1\}$, so
$\mathcal{A}(z) = \widehat{P}(Y' = 1 | Z' = z)$, but the same methods can be
used when labels belong to more than two classes. (Notation remark: training data is typically comprised of multiple conditions $C' \in \mathcal{C}'$, which may have different prevalences $P(Y' = 1 | C')$. Probabilities which don't condition on $C'$, e.g. $P(Y' = 1)$, are understood to refer to the specific combination of conditions in the observed training data). To be able to make predictions on new data from new conditions $c \in \mathcal{C}$, we assume that our classifier generalizes, as discussed above in Section \ref{sec:generalizable-background}. In particular, we assume that the covariates $Z$ satisfy the following assumptions:
\begin{itemize}
\item[(A1)] $P(Y = 1 | Z, C)$ can be estimated using the classifier $\mathcal{A}$, labeled training data $\{(Z_j', Y_j', C_j', K_j')\}_{j=1}^m$, and unlabeled test data $\{(Z_i, C_i, K_i)\}_{i=1}^n$.

\item[(A2)] $\mathcal{A}(z) = \widehat{P}(Y' = 1 | Z' = z)$ is consistent for $P(Y' = 1 | Z' = z)$.
\end{itemize}

If the features $Z$ satisfy the covariate shift assumption, (A1) is straightforward, with $\widehat{P}(Y = 1 | Z, C) = \mathcal{A}(Z)$. In other scenarios, classifier predictions $\mathcal{A}(Z)$ may need to be corrected on new data. For example, in the label shift setting, conditionwise prevalence $P(Y = 1 | C = c)$ can be estimated for each condition $c \in \mathcal{C}$ using label shift correction methods (see Appendix \ref{sec:label-shift-est}), and then $P(Y = 1 | Z, C)$ is estimated via Bayes theorem:
\begin{align}
\label{eq:label-shift-correction}
\widehat{P}(Y = 1 | Z, C) = \mathcal{A}_L(Z, C) := \frac{\frac{\widehat{P}(Y = 1 | C)}{\widehat{P}(Y' = 1)} \mathcal{A}(Z)}{\frac{\widehat{P}(Y = 1 | C)}{\widehat{P}(Y' = 1)} \mathcal{A}(Z) + \frac{1- \widehat{P}(Y = 1 | C)}{1 - \widehat{P}(Y' = 1)} (1 - \mathcal{A}(Z))},
\end{align} 
where $\mathcal{A}_L(Z, C)$ denotes the label shift-corrected predictions for
condition $C$. For the purpose of this paper, we will focus on the label shift
setting. However, we note that our work can be applied to other settings as well.

\subsection{Inference for prevalence}
\label{sec:prevalence-inference}

Our first goal is to construct a confidence interval for the conditionwise
prevalence, $P(Y = 1 | C = c)$. Given estimated probabilities $\widehat{P}(Y =
1 | Z, C)$ that are close to the true probabilities $P(Y=1 | Z, C)$, point
estimation of this quantity is straightforward: $\widehat{P}(Y = 1 | C = c) =
\frac{1}{\# \{i: C_i = c\}} \sum \limits_{i=1}^n \widehat{P}(Y_i = 1 | Z_i,
C_i) \mathbbm{1}\{C_i = c\}$. Alternatively, in the label shift setting,
$\widehat{P}(Y = 1 | C = c)$ is estimated separately by leveraging the label
shift assumption (Appendix \ref{sec:label-shift-est}). In either case, a simple binomial
confidence interval for the prevalence $P(Y=1
C = c)$ does not suffice, because $\widehat{P}(Y = 1 | C = c)$ relies on both training and test data. We therefore propose a bootstrap procedure which resamples both training and test data at each step. In particular, for bootstrap samples $s=1,...,B$ we
\begin{enumerate}
\item Resample the training data, $(Z_i'^*, Y_i'^*)$
\item Retrain the classifier, $\mathcal{A}^*$, on the bootstrap training data
\item Resample the test data $(Z_i^*, C_i^*)$
\item Re-estimate the prevalence $\widehat{P}(Y^* = 1 | C^* = c)$
\end{enumerate}

The full procedure, applied to the label shift setting, is described in Algorithm \ref{alg:bootstrap-ci-prevalence} in the Appendix. Algorithm \ref{alg:bootstrap-ci-prevalence} can also be easily modified for other forms of distributional change. For instance, in the case of covariate shift, we simply remove the label shift estimation and correction steps.

Similar bootstrap approaches are used below, for inference on $X|Y = y, C = c$. Here we make several remarks that apply to all the bootstrap procedures discussed in this paper.

\textbf{Remark:} Retraining a classifier on bootstrapped training data may be time-consuming. For certain classifiers, it may be possible to sample a new classification function without re-fitting the full model. For example, for a logistic GAM, penalizing the spline fit is equivalent to placing a prior distribution on the spline coefficients \citep{krivobokova2010simultaneous, wood2017generalized}. This results in a posterior distribution for the classifier function, given the training data, and this posterior distribution has good frequentist properties \citep{krivobokova2010simultaneous}. Then, a bootstrapped classifier $\mathcal{A}^*$ can be sampled from this posterior distribution rather than by re-fitting on bootstrapped training data. Further details are provided in the Appendix.

\textbf{Remark:} Because estimates depend on both training and test data, our bootstrap procedure resamples both training and test. This also means that coverage for the resulting bootstrap confidence intervals is defined over pairs of training and test data $\{(Z_i', Y_i', C_i', K_i')\}$, $\{(Z_i, C_i, K_i)\}$. For example, if we construct a 95\% confidence interval, in the long run 95\% of training/test pairs $\{(Z_i', Y_i', C_i', K_i')\}, \{(Z_i, C_i, K_i)\}$ will produce an interval that captures the true parameter. It is \textit{not} true that for any training set $\{(Z_i', Y_i', C_i', K_i')\}$, 95\% of future test sets $\{(Z_i, C_i, K_i)\}$ will yield a confidence interval containing the true parameter.

\textbf{Remark:} Algorithm \ref{alg:bootstrap-ci-prevalence} (in the Appendix) describes a
bootstrap procedure for confidence intervals. Here our bootstrap intervals are
first order, such as bootstrap $z$-intervals, bootstrap percentile intervals,
or bootstrap pivotal intervals. The same approach can be used for more
accuracte intervals, such as calibrated intervals, bootstrap $t$-intervals, and
BC$_a$ intervals \citep{diciccio1996bootstrap}. However, these more accurate
intervals require a second level of sampling at each bootstrap iteration, which
is likely to be computationally infeasible with classifier retraining inside
the bootstrap.

\subsection{Inference for feature distributions}
\label{sec:feature-inference}

Our second question is how the conditional distribution of a feature of
interest $X|Y = y, C = c$ differs across conditions $c$. We will focus on
confidence intervals for the class-conditional mean $\mathbb{E}[X|Y=y, C=c]$,
though other summaries of the conditional distribution could be used instead.
Given the hierarchical nature of the data, with grouping variables $C$ and $K$,
it is natural to model the conditional mean using a mixed effects model:
\begin{align}
\label{eq:me-model}
\mathbb{E}[X|Y = y, C = c, K = k] = \beta_{c,y} + b_{k},
\end{align}
where $\beta_{c,y}$ is a fixed effect and $b_{k} \sim N(0, \omega^2)$ is a random effect. 

The labels $Y$ are unobserved, but we note that
\begin{align}
\begin{split}
\mathbb{E}[X | Y = y, C = c, K = k] &= \int x f_{X|Y=y, c, k}(x) dx = \int x \frac{P(Y = y | X = x, c, k)}{P(Y = y | c,k)} f_{X|c,k}(x) dx\\ 
&= \mathbb{E}\left[X \frac{P(Y = y | X, C=c, K=k)}{P(Y = y | C=c, K=k)} \biggr\rvert C=c, K = k \right].
\end{split}
\end{align}
Therefore, we can estimate $\beta_{c,y}$ in \eqref{eq:me-model} using a weighted mixed effects model, where
\begin{align}
\label{eq:weighted-me-model}
X_i \sim N\left(\beta_{c_i, y} + b_{k_i}, \frac{\sigma_y^2}{w_{i,y}}\right), \hspace{0.5cm} b_{k} \sim N(0, \omega^2),
\end{align}
with weights $w_{i,y} = P(Y_i = y | X_i, C_i, K_i)$. The assumption of a parametric form for the random effect, which is used in bootstrapping, is necessary when we observe few levels of $K$ for each condition $C$, which is common in many scientific studies. The assumption of conditional normality for the feature of interest $X$ is used for maximum likelihood estimation (or restricted maximum likelihood estimation) of the model parameters, but is not required for inference. As we describe below, our approach to inference involves a semiparametric bootstrap which resamples residuals from the fitted model, and we see in simulations (Section \ref{sec:simulations}) that departures from conditional normality do not seem to harm the coverage of our confidence intervals.

Since the true label probabilities $P(Y = y | X, C, K)$ are unknown, we use estimated probabilities instead, yielding weights $w_{i,y} = \widehat{P}(Y_i = y | Z_i, C_i)$. This requires the assumption that the feature of interest $X$ provides no additional information about the label $Y$, after accounting for the covariates $Z$ and the condition $C$. Formally, we assume the following, which is similar to assumptions found in \cite{peters2016causal} and \cite{kuang2018stable}:
\begin{itemize}
\item[(A3)] $P(Y = y | X, Z, C, K) = P(Y = y | Z, C)$.
\end{itemize}

Fitting the model \eqref{eq:weighted-me-model} with probability weights yields
a point estimate $\widehat{\beta}_{c,y}$. To construct a confidence interval
for $\beta_{c,y}$, we bootstrap the training and test data, as in Section
\ref{sec:prevalence-inference}. When $K$ has many levels for each $C=c$, then a
hierarchical bootstrap may be employed to resample the test data. However, in
many scientific studies, $K$ often has few levels for each $C$, and so we
instead create bootstrap test data by sampling random effects and residuals. In
particular, we define residuals for each class $y \in \mathcal{Y}$ by $e_{i} =
X_i - \widehat{b}_{k_i}$, which we combine with new random effects $b_{k}^*
\sim N(0, \widehat{\omega}^2)$. In the context of label shift, for each
bootstrap sample $s = 1,...,B$ we
\begin{enumerate}
\item Resample the training data, $(Z_i'^*, Y_i'^*)$

\item Retrain the classifier, $\mathcal{A}^*$, on the bootstrap training data

\item Resample the test data:
\begin{enumerate}
\item Sample $b_{k}^* \sim N(0, \widehat{\omega}^2)$ for each group $k \in \mathcal{K}$
\item Sample $(Z_i^*, C_i^*, \mathcal{A}_L(Z_i^*, C_i^*), e_{i}^*)$ by resampling rows (to preserve any correlation between covariates $Z$ and residuals $e_{i}$)
\item Sample $Y_i^* \sim \text{Bernoulli}(A_L(Z_i^*, C_i^*))$
\item Generate new observations $X_i^*$ by $X_i^* = e_{i}^* + b_{k}^*$
\end{enumerate}
\item Calculate the label shift correction on the bootstrap training and test data, and re-fit the weighted mixed effects model
\end{enumerate}

The full details are provided in Algorithm \ref{alg:bootstrap-ci} in the Appendix. As in Section \ref{sec:prevalence-inference}, Algorithm \ref{alg:bootstrap-ci} can be modified for other forms of distributional change. For covariate shift, simply remove the label shift correction steps, and replace $\mathcal{A}_L(Z,C)$ with $\mathcal{A}(Z)$.

\textbf{Remark:} Rather than probability weights, inference may also be based
directly on binary predictions $\widehat{Y}_i \in \{0, 1\}$. The procedure is
similar, just with $\mathbb{E}[X | \widehat{Y} = 1, C = c, K = k]$. If the
classifier predictions $\mathcal{A}_L(Z, C)$ are good, we generally expect
probability weights to give better estimates than binary predictions. In
particular, if there is a relationship between $X$ and $\mathcal{A}_L(Z, C)$,
then thresholding classifier predictions to produce binary labels will lead to
biased estimates.

\textbf{Remark:} A consequence of assumption (A3) is that the random effect
$b_k$ in \eqref{eq:me-model} does not depend on the label $Y$. If random
effects are in fact label-dependent, which may be assessed with training data,
separate random effects can be estimated when fitting
\eqref{eq:weighted-me-model} and in constructing bootstrap confidence
intervals. However, label-dependent random effects violate (A3) and so may lead
to a decrease in confidence interval coverage. We investigate this further, and
suggest a potential adjustment to improve coverage in Section \ref{sec:simulations}.

\subsection{Mixture models: an alternative to probability weighting}
\label{sec:mixture-models}

Assumption (A3) states that the covariates $Z$ are sufficient for
classification. This assumption can be checked on training data, but it may be
challenging to find a subset of covariates $Z$ which satisfies both (A3) and
(A1), or even one which satisfies (A1) alone. In this case, estimating
appropriate weights for the weighted mixed effects model
\eqref{eq:weighted-me-model} may be difficult. An alternative is to recognize
that inference for the conditional feature distribution $X|Y=y, C=c$ naturally
fits a mixture model approach, with the observed distribution of $X|C=c$ being a
mixture of conditional distributions $X|Y=y, C=c$ over unobserved labels $y \in \mathcal{Y}$.

If $X|Y=y, C=c$ is assumed to follow a
parametric distribution, then maximum likelihood estimation of the model
parameters is possible. For example, we might assume a Gaussian hierarchical
mixture where
\begin{align}
\label{eq:me-mixture}
X|Y = y, C = c, K = k \ \sim \ N(\beta_{c,y} + b_{k,y}, \sigma_y^2),
\end{align}
and $b_{k,y} \sim N(0, \omega_y^2)$. This replaces assumptions (A1) - (A3) with
parametric assumptions on the conditional distribution of the feature of
interest $X$; while we use Gaussian mixture models throughout this paper, as in
\eqref{eq:me-mixture}, other parametric distributions can be chosen based on
the observed training data.
Furthermore, by removing assumption (A3), it is
straightforward to allow label-dependent random effects $b_{k,y}$ in
\eqref{eq:me-mixture}.
(\emph{Note:} the parametric assumptions for the
mixture model \eqref{eq:weighted-me-model} are much more important for
estimation than the parametric model used for mixed effect model estimation
\eqref{eq:me-mixture}. )

However, mixture models can be difficult to estimate well, particularly when
parametric assumptions are violated, and when the class distribution is
unbalanced. To assist with mixture model estimation, we can use information
from classifier predictions. In particular, we propose using a classifier to
estimate $P(Y = y | C = c)$, and then using these class proportions to improve
mixture model estimation. This approach relies on (A1) and (A2), but not (A3).

Thus we have two alternative assumptions for inference on the conditional mean
$\mathbb{E}[X|Y=y, C=c]$: that $Z$ is sufficient for classification (assumption
(A3)), or that we know a parametric form for the conditional distribution $X|Y=y,
C=c$ (as in \eqref{eq:me-mixture}). Which is assumption is more appropriate is
problem-specific. The bootstrap procedure is almost identical to
the one described in Section \ref{sec:feature-inference}, the only difference
is the model used for parameter estimation, and that residuals $e_{i,y}$ are
calculated for each class to accomdate label-dependent random effects
$b_{k,y}$. Algorithm \ref{alg:mixture-ci}, in the Appendix, describes the full procedure in
detail in the context of label shift; as before, modifications are
straightforward.

\section{Simulations}
\label{sec:simulations}

In this section, we investigate the performance of the bootstrap procedures
described in Section \ref{sec:inference}, using simulated data. All of our
methods rely on some subset of: assumpions (A1), (A2), (A3), parametric
assumptions about a feature $X$, and assumptions about the relationship between
labels and random effects. To evaluate the impact of assumptions on the
performance of our bootstrap methods, we assess coverage of bootstrap
confidence intervals when different assumptions are satisfied.

As discussed in
Section \ref{sec:prevalence-inference}, inference for label prevalence requires
assumptions (A1) and (A2), which allow label probabilities to be estimated on
new data using a subset of covariates $Z$. For inference on a feature of
interest $X$, we also require either assumption (A3) (if using the mixed
effects approach of Section \ref{sec:feature-inference}) or a known parametric
form for the class distributions (if using the parametric mixture model
approach of Section \ref{sec:mixture-models}). To evaluate performance of our
proposed confidence intervals, we consider six different simulation settings,
varying the assumptions that are satisfied. For simplicity, we consider a
single additional covariate $Z$, one test condition $C$ ($|\mathcal{C}| = 1$), and
15 nested subgroups $K$ ($|\mathcal{K}| = 15$). As in the rest of this
manuscript, we focus on label shift to illustrate our proposed procedures.

\textbf{Scenarios.} We consider three main scenarios, under which different
combinations of (A1), (A2), and (A3) are satisfied. To evaluate the sensitivity
of the mixture model approach to parametric assumptions, we use Gaussian
mixture models and generate data from both Gaussian and skewed normal
distributions for each scenario. The parameters of the skewed normal
distributions are chosen so that variance and separation between the two class
distributions are roughly equivalent to the Gaussian case. Full simulation
details are provided in Table \ref{tab:sim-settings} in the Appendix.
\begin{itemize}
\item[] \textbf{Scenario 1:} Assumptions (A1), (A2), and (A3) are satisfied. The covariate $Z$ used for classification satisfies the label shift assumption between the training and test data (A1), and a logistic spline fit is used for classification (A2). Gaussian random effects $b_k$ are generated for each $k \in \mathcal{K}$. The feature of interest $X$ is given by $X_i = Z_i + b_{k_i} + \text{noise}$, satisfying (A3).

\item[] \textbf{Scenario 2:} Assumptions (A2) and (A3) are satisfied, but (A1) is not. Though label shift methods are employed for constructing confidence intervals, as in Algorithms \ref{alg:bootstrap-ci-prevalence}, \ref{alg:bootstrap-ci}, and \ref{alg:mixture-ci} (Appendix), the conditional distribution of $Z|Y=0$ differs between training and test data. As in Scenario 1, a logistic spline fit is used for classification, and $X_i = Z_i + b_{k_i} + \text{noise}$.

\item[] \textbf{Scenario 3:} Assumptions (A1) and (A2) are satisfied, but (A3) is not. As in Scenario 1, the covariate $Z$ satisfies the label shift assumption, and a logistic spline fit is used for classification. However, $\mathbb{E}[X|Z, Y=1] \neq \mathbb{E}[X|Z, Y=0]$, which violates (A3).
\end{itemize}

\textbf{Comparisons.} For each scenario, we calculate estimates and 95\% confidence intervals for the prevalence $P(Y=1 | C = 1)$, and the class mean $\mathbb{E}[X|Y = 1, C=1]$ (recall that for the test data, we consider only one test condition, i.e. $|\mathcal{C}| = 1$). Inference for prevalence is done as in Section \ref{sec:prevalence-inference}. Inference for $\mathbb{E}[X|Y = 1, C=1]$ is done with both mixed effects models (Section \ref{sec:feature-inference}) and mixture models (Section \ref{sec:mixture-models}). Gaussian mixture models are used, and the mixing proportions are first estimated using label shift methods (Appendix \ref{sec:label-shift-est}). As the random effect $b_k$ in simulations does not depend on the class label $Y$, we modify Algorithm \ref{alg:mixture-ci} to fit a single random effect for each $k \in \mathcal{K}$, as in Algorithm \ref{alg:bootstrap-ci}. We then compare the bias of the point estimates from each method, and the observed coverage of nominal 95\% bootstrap pivotal intervals. Logistic splines were fit using the \texttt{mgcv} package \citep{mgcv} in \texttt{R}, while mixed effects models used the \texttt{lme4} package \citep{lme4}, and mixture models were implemented in \texttt{stan} using \texttt{rstan} \citep{rstan}.

\textbf{Results.} Average point estimates and confidence interval coverage are
shown in Table \ref{tab:fixed-re-results}. Inference for the prevalence
$P(Y=1|C=1)$ depends on the validity of assumptions (A1) and (A2), and Table
\ref{tab:fixed-re-results} shows that bias is small and coverage is close to
the nominal level when (A1) and (A2) are satisfied, regardless of assumption
(A3) and the parametric form of the data. The mixed effects model requires the
additional assumption (A3) in order to perform inference on the feature of
interest $X$. When (A1), (A2), and (A3) are satisfied, bias is close to 0 and
coverage is close to 95\%, and this holds for both the normal and skewed normal
distributions. However, when (A1) or (A3) are violated, the point estimates
become biased, leading to a decrease in coverage. In contrast, the Gaussian
mixture model requires (A1) and (A2), along with a feature $X$ which is conditionally normal given class $y \in \mathcal{Y}$. 

When these assumptions
are met, bias and coverage both behave well. However, departures from normality lead to
biased estimates and lower coverage. Violation of the label shift assumption
also results in poor performance, because the label shift assumption is used
to estimate mixing proportions.

\begin{table}[ht]
\centering
\tiny
\begin{tabular}{c|c|c|c|c|c|c|c}
  \toprule 
 \multirow{2}{*}{Assumptions} & \multirow{2}{*}{Normal?} & \multicolumn{2}{c|}{Prevalence} & \multicolumn{2}{c|}{Mixed Effects Model} & \multicolumn{2}{c}{Mixture Model} \\ 
 & & Mean & Coverage & Mean & Coverage & Mean & Coverage \\ 
 \midrule\multirow{2}{*}{\textbf{(A1), (A2), (A3)}} & yes & 0.4 (0.001) & 0.93 (0.015) & 2.99 (0.008) & 0.93 (0.014) & 3 (0.007) & 0.94 (0.013) \\ 
   & no & 0.4 (0.002) & 0.9 (0.018) & 3.51 (0.009) & 0.93 (0.014) & 3.86 (0.010) & 0.28 (0.026) \\ 
  \midrule\multirow{2}{*}{\textbf{(A2), (A3)}} & yes & 0.35 (0.001) & 0.07 (0.015) & 3.16 (0.008) & 0.72 (0.026) & 3.13 (0.007) & 0.8 (0.023) \\ 
   & no & 0.45 (0.002) & 0.71 (0.026) & 3.44 (0.009) & 0.91 (0.016) & 3.74 (0.009) & 0.54 (0.029) \\ 
  \midrule\multirow{2}{*}{\textbf{(A1), (A2)}} & yes & 0.4 (0.001) & 0.93 (0.015) & 3.87 (0.008) & 0.86 (0.020) & 4 (0.007) & 0.94 (0.014) \\ 
   & no & 0.4 (0.002) & 0.9 (0.018) & 4.12 (0.010) & 0.41 (0.028) & 4.68 (0.012) & 0.63 (0.028) \\ 
   \bottomrule 
\end{tabular}
\caption{Coverage of bootstrap confidence intervals in simulated data.} 
\label{tab:fixed-re-results}
\end{table}

\subsection{Label-dependent random effects}

As discussed in Section \ref{sec:feature-inference}, the weighted mixed effects
model \eqref{eq:weighted-me-model} relies on assumption (A3) to use classifier
predictions as probability weights. This assumes that the feature of
interest, $X$, adds no information to the covariates $Z$ to distinguish between
labels $Y=0$ and $Y=1$. In practice, this may be true on average across the
population but not within groups $k\in\mathcal{K}$. In particular, we may observe that the random effect $b_k$ depends on both the group $k$ and also the label $Y$, which violates (A3).

Label dependence can be accomodated in the mixed effects model by modifying
Algorithm \ref{alg:bootstrap-ci} to estimate separate separate random effects
$b_{k,y}$ for $Y=0$ and $Y=1$ within each
group $k \in \mathcal{K}$, as in Algorithm \ref{alg:mixture-ci}. Table \ref{tab:independent-re-results} shows the
results of mixed effects estimation and coverage. Here data is simulated as in
Table \ref{tab:sim-settings}, except that label-specific random effects
$b_{k,y}$ are simulated separately for each $y \in \{0,1\}$ (that is, $X_i =
Z_i + b_{k,0}\mathbbm{1}(Y_i = 0) + b_{k,1}\mathbbm{1}(Y_i = 1) +
\text{noise}$), and the model is fit by estimating each $b_{k,y}$ separately.
Comparing Table \ref{tab:independent-re-results} to Table
\ref{tab:fixed-re-results}, we see that while our point estimates perform
equivalently, label dependence for the random effects results in decreased
coverage for the confidence intervals.

This decrease in coverage arises because the probability weights in
\eqref{eq:weighted-me-model} are slightly wrong, which causes the random effect
variance $\omega_y^2$ to be underestimated. Fortunately, this issue can be
corrected by an additional variance calibration step in the bootstrap. Let
$\widehat{\omega}_y^2$, $y \in \{0,1\}$, be the initial variance estimates from
fitting weighted mixed effects models. For bootstrap samples $s=1,...,B$, we
sample bootstrap training and test data as described in Algorithm
\ref{alg:mixture-ci}, using our initial estimates $\widehat{\omega}_y^2$. For
each sample, we then calculate the observed variance estimate
$\widehat{\omega}_{y,s}^{*2}$. The same process that results in
$\widehat{\omega}_y^2$ being biased for $\omega_y^2$ will cause
$\widehat{\omega}_{y,s}^{*2}$ to be biased for $\widehat{\omega}_y^2$. Using
the $\widehat{\omega}_{y,s}^{*2}$ and $\widehat{\omega}_y^2$, we estimate a
variance correction, which we apply to $\widehat{\omega}_y^2$. The corrected
variance estimate is then used for bootstrap simulation to construct a
confidence interval. Full details are provided in the Appendix.

Table \ref{tab:independent-re-results} shows the estimates and coverage of the
variance-adjusted mixed effects bootstrap. We can see that the bias remains the
same, but adjusting the variance improves confidence interval coverage. For
comparison, we also assess the Gaussian mixture model approach with
label-dependent random effects \eqref{eq:me-mixture}. When all assumptions are
satisfied, coverage of the mixture model confidence intervals is slightly worse
than the variance-adjusted mixed effects approach, and is slightly worse than
mixture model coverage in Table \ref{tab:fixed-re-results}. This is because
allowing label-dependent random effects increases the number of quantities to
estimate in the mixture model, which makes model fitting more challenging.

\begin{table}[ht]
\centering
\tiny
\begin{tabular}{c|c|c|c|c|c|c|c}
  \toprule 
 \multirow{2}{*}{Assumptions} & \multirow{2}{*}{Normal?} & \multicolumn{2}{c|}{Mixed Effects Model} & \multicolumn{2}{c|}{\begin{tabular}[c]{@{}c@{}} Mixed Effects, \\ Variance Adjustment \end{tabular}} & \multicolumn{2}{c}{Mixture Model} \\ 
 & & Mean & Coverage & Mean & Coverage & Mean & Coverage \\ 
 \midrule\multirow{2}{*}{\textbf{(A1), (A2), (A3)}} & yes & 3 (0.007) & 0.91 (0.016) & 3 (0.007) & 0.94 (0.013) & 3.01 (0.009) & 0.89 (0.018) \\ 
   & no & 3.48 (0.007) & 0.83 (0.022) & 3.49 (0.006) & 0.93 (0.015) & 3.86 (0.008) & 0.13 (0.020) \\ 
  \midrule\multirow{2}{*}{\textbf{(A2), (A3)}} & yes & 3.16 (0.008) & 0.6 (0.028) & 3.15 (0.008) & 0.71 (0.026) & 3.12 (0.008) & 0.81 (0.022) \\ 
   & no & 3.45 (0.006) & 0.91 (0.017) & 3.42 (0.007) & 0.89 (0.018) & 3.75 (0.007) & 0.37 (0.028) \\ 
  \midrule\multirow{2}{*}{\textbf{(A1), (A2)}} & yes & 3.88 (0.007) & 0.78 (0.024) & 3.89 (0.008) & 0.84 (0.021) & 4.01 (0.008) & 0.86 (0.020) \\ 
   & no & 4.12 (0.007) & 0.18 (0.022) & 4.13 (0.008) & 0.25 (0.025) & 4.7 (0.009) & 0.5 (0.029) \\ 
   \bottomrule 
\end{tabular}
\caption{Coverage of bootstrap confidence intervals in simulated data.} 
\label{tab:independent-re-results}
\end{table}

\section{Case study: live cell microscopy data}
\label{sec:case-study}

In this section, we apply our methods for inference with classifier predictions
to a large live cell microscopy dataset that was collected with TIRF microscopy and manually labeled. This will allow
us to explore the process of assessing generalizability assumptions, engineering generalizable classifiers, and evaluating inference with unlabeled data.

\emph{Note:} Aside from the
issues we discuss in this paper, this TIRF dataset is known to have some human
label bias.  For illustrative purposes, we will ignore that label bias here and
take the human labels as ground truth.

\subsection{Data, classifier, and assumptions}

\textbf{Data.} Labeled data was collected on TIRF microscopy images under three different experimental conditions, which we will refer to as Condition 1, Condition 2, and Condition 3, denoted respectively by $C = 1, 2, 3$. The experiment recorded data for 18 different cells, with 5 cells from Condition 1, 6 from Condition 2, and 7 from Condition 3, yielding a total of 134127 events across the 18 cells, with approximately 2.7\% of the events being puffs.

Table \ref{tab:puff-count} shows the breakdown of events in each cell. Conditions 1 and 2 have similar puff rates,
while Condition 3 has a higher proportion of puffs, and furthermore it is
hypothesized that puffs for Condition 3 may have different characteristics than
puffs from Conditions 1 and 2. For this case study, we will therefore treat Conditions
1 and 2 as training data and Condition 3 as test data, allowing us to study
generalization. For statistical inference, we want to construct confidence intervals for Condition 3. To reflect the process of classifier construction and assessment, we divide Conditions 1 and 2 into training data (7 cells) and validation data (4 cells). Note that the split between training and validation data is done by cell, to better capture cell-to-cell variability.

\begin{table} 
  \small
  \centering
\begin{tabular}{|c|ccccccccccc|}
  \toprule
&\multicolumn{11}{|c|}{\textbf{Training data}} \\
\midrule
Condition & 1 & 1 & 1 & 1 & 1 & 2 & 2 & 2 & 2 & 2 & 2\\
Cell & 1-1 & 1-2 & 1-3 & 1-4 & 1-5 & 2-1 & 2-2 & 2-3 & 2-4 & 2-5 & 2-6 \\
\# Puffs & 71 & 254 & 78 & 35 & 118 & 62 & 108 & 37 & 34 & 72 & 16 \\
Puff prevalence & 0.012 & 0.024 & 0.011 & 0.005 & 0.020 & 0.007 & 0.011 & 0.004 & 0.010 & 0.009 & 0.003 \\

\bottomrule
\end{tabular}\\
\vspace{1em}
\begin{tabular}{|c|ccccccc|}
  \toprule
& \multicolumn{7}{|c|}{\textbf{Test data}} \\
\midrule
Condition & 3 & 3 & 3 & 3 & 3 & 3 & 3 \\
Cell & 3-1 & 3-2 & 3-3 & 3-4 & 3-5 & 3-6 & 3-7  \\
\# Puffs & 147 & 488 & 509 & 482 & 266 & 604 & 183  \\
Puff prevalence & 0.024 & 0.052 & 0.050 & 0.062 & 0.040 & 0.075 & 0.049  \\
   \bottomrule
\end{tabular}

\caption{Breakdown of events in each cell of the TIRF microscopy data, showing the number of puffs and the prevalence.}
\label{tab:puff-count}
\end{table}

\textbf{Features.}  We will investigate the hypothesis that the pattern of
diffusion into the cell membrane after vesicle fusion differs between conditions.
To capture these diffusion differences, we create a feature called
$Smoothness$, which we expect may vary across conditions.  The feature is
construted using a functional PCA approach on the raw images (see Appendix) to
capture a smoothness aspect of the diffusion. In the notation from Section \ref{sec:intro}, $Smoothness$ is our feature of interest $X$.

We also construct a set of features $Z$ for classification. These are carefully
designed to capture the fundamental characteristics of puffs shown in Figure
\ref{fig:tirf}, and are also created from a functional PCA representation of
the images (see Appendix).  Because we expect common geometric characteristics of puffs
to be preserved across conditions, but that the rate of puffs will differ, we will
aim for generalizable prediction by constructing a set of features---and
therefore a classifier---that obey the label shift assumption.  We construct
these by comparing the
distributions of our designed features across conditions, selecting those
features for
which the label shift assumption appears to hold in exploratory data analysis.
These are:
\begin{itemize}
\item $IntensityRatio$: the event's maximum intensity / minimum intensity ($\Delta_f$)
\item $SNR$: a measure of the signal-to-noise in the event
\item $ConvexArea$ and $ConvexPerimeter$: measures of the amount of diffusion in the event's intensity over time
\item $Noise$: a measure of randomness in the event's time series
\end{itemize}

\begin{figure}
\centering
\includegraphics[scale=0.45]{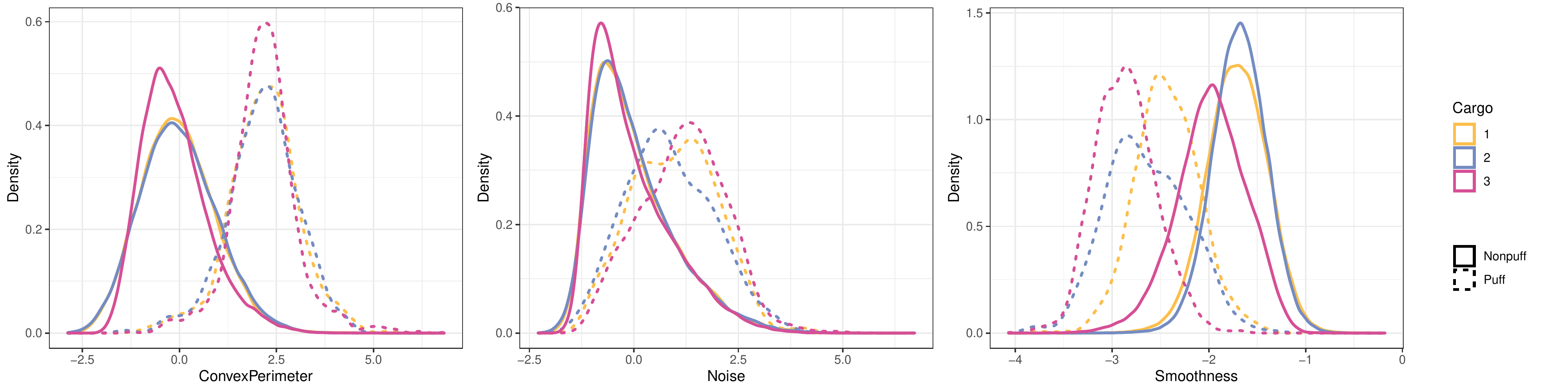}
\caption{Distribution of three features calculated on the TIRF events, broken down by condition and puff/nonpuff. The first two features show label shift between the different conditions, and will be used to construct a classifier. The distribution of the third feature, $Smoothness$, changes between conditions. Inference on $Smoothness$ in new conditions is of interest, but the feature is not included in the classifier because of the change in distribution.}
\label{fig:feature-distributions}
\end{figure}

For example, Figure \ref{fig:feature-distributions} shows the distribution of
$ConvexArea$ and $ConvexPerimeter$; as the feature distribution within each
class (puff and nonpuff) is the same between conditions, the only difference is one
of label shift, which supports (A1). In contrast, the distribution of $X$
($Smoothness$) does not satisfy label shift, as expected (Figure
\ref{fig:feature-distributions}). 

To evaluate assumptions (A2) and (A3), we first need to construct a classifier.

\begin{figure}
\includegraphics[scale=0.37]{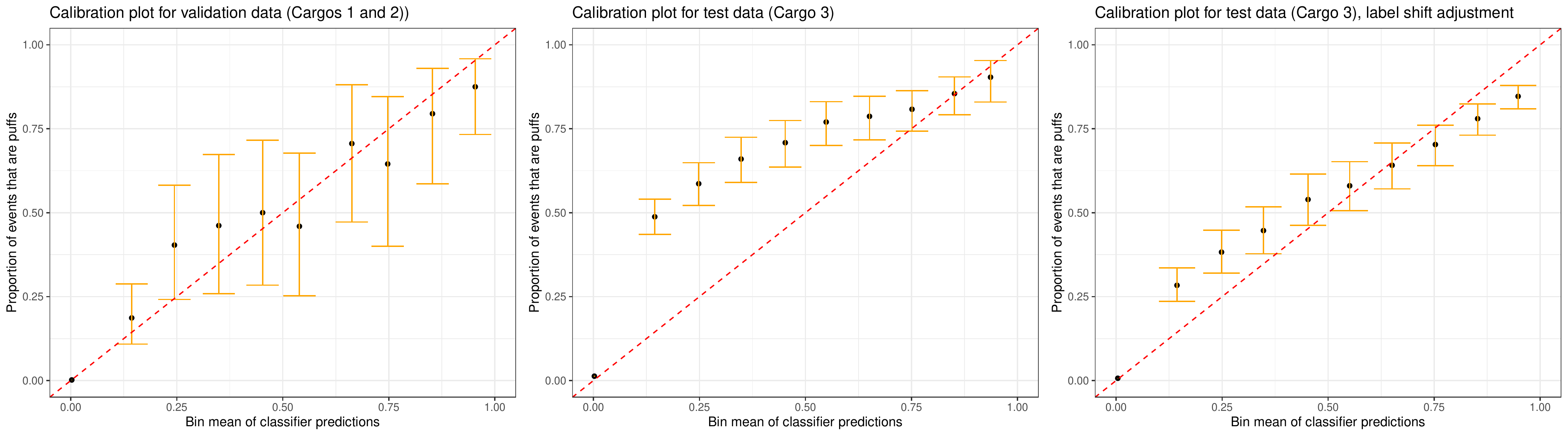}
\caption{Calibration plots for the classifier on validation (left panel) and test data (middle and right panels). Because TfR cells have a higher proportion of puffs (Table \ref{tab:puff-count}), classifier predictions are mis-calibrated (middle panel), but they can be corrected with a label shift adjustment (right panel).}
\label{fig:cal-plots}
\end{figure}

\textbf{Classifier.} A logistic GAM \citep{hastie1990generalized,
wood2017generalized} is trained on the seven training cells, using features $Z
= $ $IntensityRatio$, $SNR$, $ConvexArea$, $ConvexPerimeter$, and $Noise$. We
choose a logistic GAM because it provides a balance between simplicity and
flexibility, and because sampling from the posterior (see Appendix) makes our
bootstrap procedures much more efficient. Performance of the classifier is
assessed on the four validation cells, and the classifier is then applied to
the test data. Use of validation data allows us to assess performance of
classifier predictions on new data from the training distribution, which
provides a benchmark for performace on new data from a different distribution.
Figure \ref{fig:cal-plots} shows calibration plots for the predicted
probabilities on validation and test data. As expected, the predicted
probabilities appear to be  calibrated on the validation data from Conditions 1 and
2 (which supports (A2)), but are mis-calibrated---due to label shift---on the
test data from Condition 3. Furthermore, because the label shift assumption is
appropriate for our classifier features (Figure
\ref{fig:feature-distributions}), $P(Y = 1 | C = 3)$ can be estimated as in
the Appendix \ref{sec:label-shift-est}. The resulting estimate is 0.043 using the method from \cite{lipton2018detecting}, and 0.050 using the fixed point method based on \cite{saerens2002adjusting} (Section
\ref{sec:caseinference}), compared to the
true sample prevalence of 0.051. Using this estimate, we correct the classifier predictions as in
\eqref{eq:label-shift-correction} with Bayes theorem, and Figure
\ref{fig:cal-plots} shows the corrected predictions are much better calibrated,
which again supports (A1).

\begin{figure}
\centering
\includegraphics[scale=0.4]{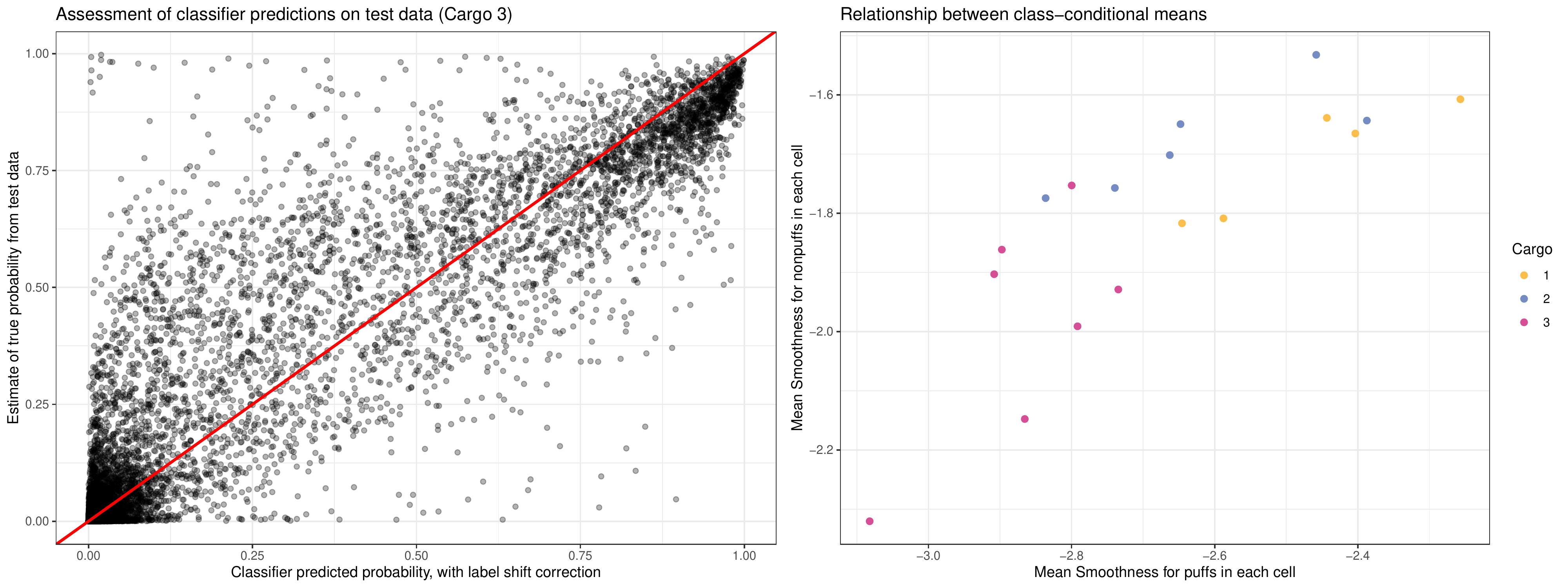}
\caption{\underline{Left}: $\widehat{P}(Y_i = 1 | X_i, Z_i)$ vs. $\mathcal{A}_L(Z_i, C_i)$ on test data. \underline{Right}: Relationship between the class-conditional means $\mathbb{E}[X|Y=1, C=c, K=k]$ and $\mathbb{E}[X|Y=0, C=c, K=k]$ for each cell condition $c$ and cell $k$.}
\label{fig:prediction-assessment}
\end{figure}

Finally, to conduct inference about $X|Y=y, C=c$, we require (A3) to hold. To
assess (A3), we compare $\mathcal{A}_L(Z, C)$ from
\eqref{eq:label-shift-correction} to an estimate of $P(Y=1|X, Z, C)$ from
regression on the test data. The resulting plot is shown in Figure
\ref{fig:prediction-assessment}, which suggests that while not perfect, the
corrected predictions $\mathcal{A}_L(Z, C)$ are a good estimate of the true
probability $P(Y=1|X, Z, C)$. As discussed in Section \ref{sec:inference} and
Section \ref{sec:simulations}, a consequence of (A3) is that the per-cell
random effect $b_k$ should not depend on the label $Y$. Figure
\ref{fig:prediction-assessment} also shows the relationship between
$\mathbb{E}[X|Y = 1, C=c, K=k]$ and $\mathbb{E}[X|Y=0, C=c, K=k]$. As the
relationship is roughly linear with a slope of approximately 1, the assumption
of label-independent random effects is not unreasonable, but confidence
intervals can also be constructed that model label-dependent random effects if
desired, as described in Section \ref{sec:simulations}.

\subsection{Inference with classifier predictions}
\label{sec:caseinference}

\textbf{Inference for prevalence.} We begin with inference for $P(Y=1|C=3)$. Our point estimate, from label shift estimation, was 0.050 using the fixed point method (see \cite{saerens2002adjusting} and Appendix \ref{sec:label-shift-est}) and 0.043 using the method from \cite{lipton2018detecting}. Using the procedure described in Algorithm \ref{alg:bootstrap-ci-prevalence}, we construct a 95\% confidence interval for $P(Y=1|C=3)$ with each label shift method, which are respectively (0.047, 0.055) and (0.039, 0.049). Both label shift estimates are a marked improvement on the point estimate from uncorrected classifier probabilities, which is 0.029.

To assess coverage of this interval, we perform a simulation using the real TIRF microscopy data. We simulate new training and test sets by resampling from the original training and test data, using Algorithm \ref{alg:bootstrap-ci-prevalence} to calculate a bootstrap confidence interval for the prevalence $P(Y=1|C=3)$ in each simulation. The simulated training data ($C = 1,2$) has a prevalence of 0.01, while the simulated test data ($C = 3$) has a prevalence of 0.05. Using the fixed point method (Appendix \ref{sec:label-shift-est} and \cite{saerens2002adjusting}), nominal 95\% bootstrap pivotal intervals have a coverage of 95\% in simulations, while coverage using the discretization method \citep{lipton2018detecting} is about 19\%. The lower coverage using the discretization method is due to bias, which may result because the prevalences are close to 0, or because the label shift assumption is not perfectly satisfied.

\textbf{Inference for feature distributions.} Next, we are interested in
constructing a confidence interval for $\mathbb{E}[X|Y=1, C=3]$, where $X$ is
diffusion $Smoothness$. We first examine estimates of mean puff $Smoothness$ in
each test cell, using classifier predictions. Table \ref{tab:test-means} shows
estimates using both probability weights and binary predictions from
thresholding, and compares the classifier with and without label shift
correction. We can see that the estimated means are close to the true sample
means in each cell, and as expected the label shift correction produces better
estimates. The weighted mean generally does better than the thresholding-based
mean, which is biased because classifier predictions are negatively associated
with $Smoothness$.

\begin{table}[]
\small
\centering
\begin{tabular}{c|cccccc}
\hline
Cell     & \begin{tabular}[c]{@{}c@{}}Puff\\ mean\end{tabular} & \begin{tabular}[c]{@{}c@{}}Weighted mean,\\ raw\end{tabular} & \begin{tabular}[c]{@{}c@{}}Weighted mean,\\ label shift\end{tabular} & \begin{tabular}[c]{@{}c@{}}Threshold mean,\\ raw\end{tabular} & \begin{tabular}[c]{@{}c@{}}Threshold mean,\\ label shift\end{tabular} & \begin{tabular}[c]{@{}c@{}}Nonpuff\\ mean\end{tabular} \\ \hline
3-1 & -3.08                                               & \textbf{-3.09}                                               & -3.01                                                                & -3.19                                                         & -3.11                                                                 & -2.32                                                  \\
3-2 & -2.90                                               & -2.97                                                        & \textbf{-2.84}                                                       & -3.10                                                         & -2.99                                                                 & -1.86                                                  \\
3-3 & -2.91                                               & -2.95                                                        & -2.86                                                                & -3.05                                                         & \textbf{-2.94}                                                        & -1.90                                                  \\
3-4 & -2.87                                               & -2.99                                                        & \textbf{-2.92}                                                       & -3.10                                                         & -2.99                                                                 & -2.15                                                  \\
3-5 & -2.80                                               & -2.92                                                        & \textbf{-2.82}                                                       & -3.06                                                         & -2.94                                                                 & -1.75                                                  \\
3-6 & -2.79                                               & -2.92                                                        & \textbf{-2.85}                                                       & -3.04                                                         & -2.93                                                                 & -1.99                                                  \\
3-7 & -2.73                                               & -2.78                                                        & -2.67                                                                & -2.90                                                         & \textbf{-2.77}                                                        & -1.93                                                  \\
\hline
Combined         & -2.87                                               & -2.95                                                        & \textbf{-2.85}                                                       & -3.06                                                         & -2.95                                                                 & -1.99         \\
\hline                                         
\end{tabular}
\caption{Performance of estimates with classifier predictions on test data (Condition 3).}
\label{tab:test-means}
\end{table}

The point estimate for $\mathbb{E}[X|Y=1, C=3]$, using the weighted mixed effects model \eqref{eq:weighted-me-model}, is -2.85. Using the procedure described in Algorithm \ref{alg:bootstrap-ci}, we construct a confidence interval, which is (-3.00, -2.73). To assess coverage of our confidence interval on TIRF microscopy data, we simulate new training and test sets from the real data. Training data is sampled by bootstrapping from the original training data, and test data is simulated by adding a per-cell random effect to $Smoothness$ in bootstrap samples from the original test data. For each simulated training and test pair, the procedure in Algorithm \ref{alg:bootstrap-ci} is used to construct a confidence interval, and coverage is assessed across training/test pairs. In these simulations, nominal 95\% bootstrap pivotal intervals have a coverage of about 90\%, with similar numbers for other first-order intervals like bootstrap percentile intervals. This is close to the coverage seen in our simulations in Section \ref{sec:simulations}.

Using a modification of Algorithm \ref{alg:bootstrap-ci} to accomodate label-dependent random effects, a 95\% bootstrap confidence interval for $\mathbb{E}[X|Y=1, C=3]$ in the test data is (-2.93, -2.78). In simulations with label-dependent random effects, these nominal 95\% intervals again have a coverage of about 90\%. The difference in width between the two confidence intervals likely results from different estimates of the random effect variance: as the nonpuff cell means vary slightly more than the puff cell means in the observed data, bootstrap puff data varies less when we allow label-dependent random effects. 

\textbf{Mixture models.} As an alternative to probability weighting, we employ the mixture model approach described in Section \ref{sec:mixture-models}. We use a two-component Gaussian mixture, fitting the hierarchical mixture described in \eqref{eq:me-mixture}. To improve estimation of the mixture model, we pre-specify the prevalence of puffs in the test data, using the label shift estimate that we calculated above. Our point estimate for $\mathbb{E}[X|Y=1,C=3]$ is then $\widehat{\beta}_{3,1} =
-2.95$, and the confidence interval from Algorithm \ref{alg:mixture-ci} is
(-2.99, -2.85). To assess coverage, we simulate from training/test pairs as with the
mixed effects model above. Even though pre-specifying the puff prevalence improves the mixture estimation, the nominal 95\% bootstrap
percentile intervals have a coverage of approximately 15\%. The poor coverage
here is due largely to bias resulting from skewness in the data, which matches
the simulation results for the skewed normal distribution in Section
\ref{sec:simulations}. We experimented with other parametric models beyond a
Gaussian mixture, but estimation remained challenging. Figures showing the fitted Gaussian mixture model in each cell (with and without pre-specifying puff prevalence) can be found in the Appendix, in addition to the estimated means in each cell.

\section{Discussion}

Scientific studies often require painstakingly labeling large volumes of
unlabeled data, causing labeling to be a  key limiting factor in data
analysis. If data can be automatically labeled with predictive models, 
manual labeling costs can be dramatically reduced and much higher through-put
science can be enabled. However, rigorous scientific analysis with predicted
labels requires generalizable classifier predictions for valid inference.

We have described methods for valid inference two common downstream targets of
inference: the label prevalence $P(Y = y |
C = c)$ and class-conditional feature means $\mathbb{E}[X|Y = y, C = c]$.
Inspired by our motivating example from TIRF microscopy, we focus on the case where
the classifier used for automatic labeling is trained on data that differs in
distribution from the new, unlabeled data.  As this dataset shift may
prevent classifiers from generalizing to the new data, and therefore prevent valid
statistical inference, we rely on identification of a subset of features that
enable construction of a generalizable classifier. These features can be designed from
training data, and a variety of methods exist to construct features which
satisfy the covariate shift assumption \citep{peters2016causal,
magliacane2017domain, kuang2018stable, rojas2018invariant,
subbaswamy2019preventing}. In our TIRF microscopy case study, a label shift
assumption is more appropriate than covariate shift, and we show that
exploratory data analysis and careful feature engineering can construct
generalizable covariates.

While we focus on the label shift setting in our algorithms, simulations, and
case study, our methods can be easily modified for other types of dataset
shift. Furthermore, our methods are designed to accomodate a flexible
hierarchical data structure, which is common to scientific experiments in which
multiple repetitions of the experiment are performed for each experimental
condition. Through simulations and a case study with TIRF microscopy data, we
show that the tools presented in this manuscript allow statistical inference
with unlabeled, classifier-scored data, and that a generalizable classifier is
crucial for valid analysis.

For inference with the class-conditional means $\mathbb{E}[X|Y=y, C=c]$, we
describe two approaches for using classifier predictions for estimation and
confidence intervals: weighted mixed effects models that use classifier
predictions as probability weights, and hierarchical mixture models that use
classification to estimate mixing proportions. The specific assumptions
required by these two approaches are detailed in Section
\ref{sec:feature-inference} and Section \ref{sec:mixture-models}. As seen in
simulations and the TIRF microscopy case study, the weighted mixed effects
method is particularly sensitive to violations of assumption (A3), that the
feature of interest $X$ does not help distinguish between the classes $y \in
\mathcal{Y}$ once we condition on the classifier features $Z$. On the other
hand, the mixture model method is particularly sensitive to departures from the
assumed parametric class distributions. Which assumption is more appropriate is
problem dependent; fortunately, though assumptions typically cannot be checked
on test data, they can be assessed on training data. As a result, collecting a
large and diverse training set, with observations from many domains, is an
important part of data analysis.

The methods we propose in this manuscript allow valid statistical inference under appropriate assumptions, but there are still limitations to inference on unlabeled data. Even if classifier generalizability assumptions hold for inference on some features of interest, they may fail for others. In other cases, it may be impossible to construct features which satisfy the necessary conditions. Interpretation of inference results is also more nuanced: since confidence intervals must account for variability in the training data, coverage applies to joint training/test pairs, rather than to all new test datasets. Even more subtly, we often rely on humans to provide ``ground truth'' labels for classifier training; however, if researchers are accustomed to labeling observations under only some experimental conditions, then manual labels may suffer the same generalizability problems as automated predictions. While the methods in this paper may not help generalize human predictions, we believe that the explicit assumptions discussed above can still help researchers reflect on their own manual labeling system.

\section*{Acknowledgments}

The authors would like to thank Hao Chen, Jennifer M. Kunselman, and Stephanie E. Crilly for advice and assistance, and Alexandra Chouldechova for computational resources. This work was supported by NSF DMS1613202 (CE and MG), NIH K12GM081266 (ZYW), and NIH GM117425 (MAP).

\bibliographystyle{apalike}
\bibliography{references}

\begin{thebibliography}{}

\bibitem[Argyriou et~al., 2008]{argyriou2008convex}
Argyriou, A., Evgeniou, T., and Pontil, M. (2008).
\newblock Convex multi-task feature learning.
\newblock {\em Machine learning}, 73(3):243--272.

\bibitem[Argyriou et~al., 2007]{argyriou2007spectral}
Argyriou, A., Micchelli, C.~A., Pontil, M., and Ying, Y. (2007).
\newblock A spectral regularization framework for multi-task structure
  learning.
\newblock In {\em NIPS}, volume 1290, page 1296. Citeseer.

\bibitem[Axelrod, 1981]{axelrod1981tirf}
Axelrod, D. (1981).
\newblock {Cell-substrate contacts illuminated by total internal reflection
  fluorescence.}
\newblock {\em The Journal of Cell Biology}, 89(1):141--145.

\bibitem[Bates et~al., 2015]{lme4}
Bates, D., M{\"a}chler, M., Bolker, B., and Walker, S. (2015).
\newblock Fitting linear mixed-effects models using {lme4}.
\newblock {\em Journal of Statistical Software}, 67(1):1--48.

\bibitem[Bickel et~al., 2009]{bickel2009discriminative}
Bickel, S., Br{\"u}ckner, M., and Scheffer, T. (2009).
\newblock Discriminative learning under covariate shift.
\newblock {\em Journal of Machine Learning Research}, 10(9).

\bibitem[Bohannon et~al., 2017]{bohannon2017exocytosis}
Bohannon, K.~P., Bittner, M.~A., Lawrence, D.~A., Axelrod, D., and Holz, R.~W.
  (2017).
\newblock {Slow fusion pore expansion creates a unique reaction chamber for
  co-packaged cargo}.
\newblock {\em The Journal of general physiology}, 149(10):921--934.

\bibitem[Bowman et~al., 2015]{bowman2015cell}
Bowman, S.~L., Soohoo, A.~L., Shiwarski, D.~J., Schulz, S., Pradhan, A.~A., and
  Puthenveedu, M.~A. (2015).
\newblock Cell-autonomous regulation of mu-opioid receptor recycling by
  substance p.
\newblock {\em Cell reports}, 10(11):1925--1936.

\bibitem[Caicedo et~al., 2019]{caicedo2019nucleus}
Caicedo, J.~C., Goodman, A., Karhohs, K.~W., Cimini, B.~A., Ackerman, J.,
  Haghighi, M., Heng, C., Becker, T., Doan, M., McQuin, C., et~al. (2019).
\newblock Nucleus segmentation across imaging experiments: the 2018 data
  science bowl.
\newblock {\em Nature methods}, 16(12):1247--1253.

\bibitem[Christiansen et~al., 2018]{christiansen2018silico}
Christiansen, E.~M., Yang, S.~J., Ando, D.~M., Javaherian, A., Skibinski, G.,
  Lipnick, S., Mount, E., O’Neil, A., Shah, K., Lee, A.~K., et~al. (2018).
\newblock In silico labeling: predicting fluorescent labels in unlabeled
  images.
\newblock {\em Cell}, 173(3):792--803.

\bibitem[DiCiccio et~al., 1996]{diciccio1996bootstrap}
DiCiccio, T.~J., Efron, B., et~al. (1996).
\newblock Bootstrap confidence intervals.
\newblock {\em Statistical science}, 11(3):189--228.

\bibitem[Garg et~al., 2020]{garg2020unified}
Garg, S., Wu, Y., Balakrishnan, S., and Lipton, Z.~C. (2020).
\newblock A unified view of label shift estimation.
\newblock {\em arXiv preprint arXiv:2003.07554}.

\bibitem[Gretton et~al., 2009]{gretton2009covariate}
Gretton, A., Smola, A., Huang, J., Schmittfull, M., Borgwardt, K., and
  Sch{\"o}lkopf, B. (2009).
\newblock Covariate shift by kernel mean matching.
\newblock {\em Dataset shift in machine learning}, 3(4):5.

\bibitem[Hastie and Tibshirani, 1990]{hastie1990generalized}
Hastie, T.~J. and Tibshirani, R.~J. (1990).
\newblock {\em Generalized additive models}, volume~43.
\newblock CRC press.

\bibitem[Kou et~al., 2019]{kou2019exocytosis}
Kou, Z.~W., Mo, J.~L., Wu, K.~W., Qiu, M.~H., Huang, Y.~L., Tao, F., Lei, Y.,
  Lv, L.~L., and Sun, F.~Y. (2019).
\newblock {Vascular endothelial growth factor increases the function of
  calcium-impermeable AMPA receptor GluA2 subunit in astrocytes via activation
  of protein kinase C signaling pathway}.
\newblock {\em Glia}, 67(7):1344--1358.

\bibitem[Kraus et~al., 2017]{kraus2017automated}
Kraus, O.~Z., Grys, B.~T., Ba, J., Chong, Y., Frey, B.~J., Boone, C., and
  Andrews, B.~J. (2017).
\newblock Automated analysis of high-content microscopy data with deep
  learning.
\newblock {\em Molecular systems biology}, 13(4):924.

\bibitem[Krivobokova et~al., 2010]{krivobokova2010simultaneous}
Krivobokova, T., Kneib, T., and Claeskens, G. (2010).
\newblock Simultaneous confidence bands for penalized spline estimators.
\newblock {\em Journal of the American Statistical Association},
  105(490):852--863.

\bibitem[Kuang et~al., 2018]{kuang2018stable}
Kuang, K., Cui, P., Athey, S., Xiong, R., and Li, B. (2018).
\newblock Stable prediction across unknown environments.
\newblock In {\em Proceedings of the 24th ACM SIGKDD International Conference
  on Knowledge Discovery \& Data Mining}, pages 1617--1626.

\bibitem[Lipton et~al., 2018]{lipton2018detecting}
Lipton, Z.~C., Wang, Y.-X., and Smola, A. (2018).
\newblock Detecting and correcting for label shift with black box predictors.
\newblock {\em arXiv preprint arXiv:1802.03916}.

\bibitem[Logan et~al., 2017]{logan2017exocytosis}
Logan, T., Bendor, J., Toupin, C., Thorn, K., and Edwards, R.~H. (2017).
\newblock {$\alpha$-Synuclein promotes dilation of the exocytotic fusion pore}.
\newblock {\em Nature Neuroscience}, 20(5):681--689.

\bibitem[Magliacane et~al., 2017]{magliacane2017domain}
Magliacane, S., van Ommen, T., Claassen, T., Bongers, S., Versteeg, P., and
  Mooij, J.~M. (2017).
\newblock Domain adaptation by using causal inference to predict invariant
  conditional distributions.
\newblock {\em arXiv preprint arXiv:1707.06422}.

\bibitem[Maretto et~al., 2020]{maretto2020spatio}
Maretto, R.~V., Fonseca, L.~M., Jacobs, N., K{\"o}rting, T.~S., Bendini, H.~N.,
  and Parente, L.~L. (2020).
\newblock Spatio-temporal deep learning approach to map deforestation in amazon
  rainforest.
\newblock {\em IEEE Geoscience and Remote Sensing Letters}.

\bibitem[Norouzzadeh et~al., 2018]{norouzzadeh2018automatically}
Norouzzadeh, M.~S., Nguyen, A., Kosmala, M., Swanson, A., Palmer, M.~S.,
  Packer, C., and Clune, J. (2018).
\newblock Automatically identifying, counting, and describing wild animals in
  camera-trap images with deep learning.
\newblock {\em Proceedings of the National Academy of Sciences},
  115(25):E5716--E5725.

\bibitem[Pan and Yang, 2009]{pan2009survey}
Pan, S.~J. and Yang, Q. (2009).
\newblock A survey on transfer learning.
\newblock {\em IEEE Transactions on knowledge and data engineering},
  22(10):1345--1359.

\bibitem[Peters et~al., 2016]{peters2016causal}
Peters, J., B{\"u}hlmann, P., and Meinshausen, N. (2016).
\newblock Causal inference by using invariant prediction: identification and
  confidence intervals.
\newblock {\em Journal of the Royal Statistical Society. Series B (Statistical
  Methodology)}, pages 947--1012.

\bibitem[Pippig et~al., 1995]{pippig1995receptors}
Pippig, S., Andexinger, S., and Lohse, M.~J. (1995).
\newblock {Sequestration and recycling of beta 2-adrenergic receptors permit
  receptor resensitization.}
\newblock {\em Molecular Pharmacology}, 47(4):666--676.

\bibitem[Qui{\~n}onero-Candela et~al., 2009]{quinonero2009dataset}
Qui{\~n}onero-Candela, J., Sugiyama, M., Lawrence, N.~D., and Schwaighofer, A.
  (2009).
\newblock {\em Dataset shift in machine learning}.
\newblock Mit Press.

\bibitem[Rammer and Seidl, 2019]{rammer2019harnessing}
Rammer, W. and Seidl, R. (2019).
\newblock Harnessing deep learning in ecology: An example predicting bark
  beetle outbreaks.
\newblock {\em Frontiers in plant science}, 10:1327.

\bibitem[Ramsay and Silverman, 2005]{ramsay2005fda}
Ramsay, J. and Silverman, B. (2005).
\newblock {\em Functional data analysis}.
\newblock Springer.

\bibitem[Rappoport et~al., 2003]{rappoport2003tirf}
Rappoport, J.~Z., Taha, B.~W., Lemeer, S., Benmerah, A., and Simon, S.~M.
  (2003).
\newblock {The AP-2 Complex Is Excluded from the Dynamic Population of Plasma
  Membrane-associated Clathrin}.
\newblock {\em Journal of Biological Chemistry}, 278(48):47357--47360.

\bibitem[Rojas-Carulla et~al., 2018]{rojas2018invariant}
Rojas-Carulla, M., Sch{\"o}lkopf, B., Turner, R., and Peters, J. (2018).
\newblock Invariant models for causal transfer learning.
\newblock {\em The Journal of Machine Learning Research}, 19(1):1309--1342.

\bibitem[Saerens et~al., 2002]{saerens2002adjusting}
Saerens, M., Latinne, P., and Decaestecker, C. (2002).
\newblock Adjusting the outputs of a classifier to new a priori probabilities:
  a simple procedure.
\newblock {\em Neural computation}, 14(1):21--41.

\bibitem[Sankaranarayanan et~al., 2000]{sankaranarayanan2000gfp}
Sankaranarayanan, S., De~Angelis, D., Rothman, J.~E., and Ryan, T.~a. (2000).
\newblock {The Use of pHluorins for Optical Measurements of Presynaptic
  Activity}.
\newblock {\em Biophysical Journal}, 79(4):2199--2208.

\bibitem[Shimodaira, 2000]{shimodaira2000improving}
Shimodaira, H. (2000).
\newblock Improving predictive inference under covariate shift by weighting the
  log-likelihood function.
\newblock {\em Journal of statistical planning and inference}, 90(2):227--244.

\bibitem[{Stan Development Team}, 2020]{rstan}
{Stan Development Team} (2020).
\newblock {RStan}: the {R} interface to {Stan}.
\newblock R package version 2.21.2.

\bibitem[Storkey, 2009]{storkey2009training}
Storkey, A. (2009).
\newblock When training and test sets are different: characterizing learning
  transfer.
\newblock {\em Dataset shift in machine learning}, pages 3--28.

\bibitem[Subbaswamy et~al., 2019]{subbaswamy2019preventing}
Subbaswamy, A., Schulam, P., and Saria, S. (2019).
\newblock Preventing failures due to dataset shift: Learning predictive models
  that transport.
\newblock In {\em The 22nd International Conference on Artificial Intelligence
  and Statistics}, pages 3118--3127. PMLR.

\bibitem[Sugiyama et~al., 2008]{sugiyama2008direct}
Sugiyama, M., Suzuki, T., Nakajima, S., Kashima, H., von B{\"u}nau, P., and
  Kawanabe, M. (2008).
\newblock Direct importance estimation for covariate shift adaptation.
\newblock {\em Annals of the Institute of Statistical Mathematics},
  60(4):699--746.

\bibitem[Swanson et~al., 2015]{swanson2015snapshot}
Swanson, A., Kosmala, M., Lintott, C., Simpson, R., Smith, A., and Packer, C.
  (2015).
\newblock Snapshot serengeti, high-frequency annotated camera trap images of 40
  mammalian species in an african savanna.
\newblock {\em Scientific data}, 2(1):1--14.

\bibitem[Way et~al., 2021]{way2021predicting}
Way, G.~P., Kost-Alimova, M., Shibue, T., Harrington, W.~F., Gill, S.,
  Piccioni, F., Becker, T., Shafqat-Abbasi, H., Hahn, W.~C., Carpenter, A.~E.,
  et~al. (2021).
\newblock Predicting cell health phenotypes using image-based morphology
  profiling.
\newblock {\em Molecular Biology of the Cell}, 32(9):995--1005.

\bibitem[Wood, 2011]{mgcv}
Wood, S.~N. (2011).
\newblock Fast stable restricted maximum likelihood and marginal likelihood
  estimation of semiparametric generalized linear models.
\newblock {\em Journal of the Royal Statistical Society (B)}, 73(1):3--36.

\bibitem[Wood, 2017]{wood2017generalized}
Wood, S.~N. (2017).
\newblock {\em Generalized additive models: an introduction with R}.
\newblock CRC press.

\bibitem[Yu et~al., 1993]{yu1993receptors}
Yu, S.~S., Lefkowitz, R.~J., and Hausdorff, W.~P. (1993).
\newblock {Beta-adrenergic receptor sequestration. A potential mechanism of
  receptor resensitization.}
\newblock {\em Journal of Biological Chemistry}, 268(1):337--341.

\bibitem[Yudowski et~al., 2006]{yudowski2006distinct}
Yudowski, G.~A., Puthenveedu, M.~A., and von Zastrow, M. (2006).
\newblock Distinct modes of regulated receptor insertion to the somatodendritic
  plasma membrane.
\newblock {\em Nature neuroscience}, 9(5):622--627.

\end{thebibliography}

\appendix

\section{Appendix}

\subsection{Label shift estimation}
\label{sec:label-shift-est}

Let $\{(Z_i', Y_i', C_i')\}_{i=1}^m$ denote a set of labeled training data, and $\{(Z_j, C_j)\}_{j=1}^n$ a set of unlabeled test data with unobserved labels $Y_j \in \{0,1\}$. Let $\mathcal{A}$ denote a classifier fit on the training data, with $\mathcal{A}(z) = \widehat{P}(Y_i' = 1 | Z_i' = z)$. Under the label shift assumption, $P(Y = 1 | Z = z, C = c) \neq P(Y' = 1 | Z' = z, C' = c')$, and classifier predictions $\mathcal{A}(z)$ must be corrected to $\mathcal{A}_L(z,c)$ by equation \eqref{eq:label-shift-correction}.

This requires estimating the prevalence $P(Y = 1 | C = c)$ for each $c \in \mathcal{C}$, which is challenging when labels are unobserved. Fortunately, the test prevalence $P(Y = 1 | C = c)$ can be estimated under the label shift assumption. Here we summarize two different approaches, which in our experience are simple but generally reliable for label shift estimation.

\subsubsection{Discretization method}

The discretization approach is due to \cite{lipton2018detecting}, and we summarize the method in Algorithm \ref{alg:lipton-ls}. Their approach relies on discretizing the predicted probabilities $\mathcal{A}(Z)$, and recognizing that under label shift $\mathcal{A}(Z) \ \indep \ C | Y$.

\begin{algorithm}
\small
\KwData{Labeled training data $\{(Z_i', Y_i', C_i')\}_{i=1}^m$ \newline
Unlabeled test data $\{(Z_j, C_j)\}_{j=1}^n$ \newline
Classifier $\mathcal{A}$ with $\mathcal{A}(z) = \widehat{P}(Y_i' = 1 | Z_i' = z)$}
 \SetKwInOut{Input}{Input}
    \SetKwInOut{Output}{Output}

    \Input{Discretization threshold $h \in (0, 1)$}
    \Output{Estimated prevalence $\widehat{P}(Y=1|C = c)$ for each $c \in \mathcal{C}$}
    \Init{}{
Calculate discretized predictions on training data: $\widehat{Y}_i' = \mathbbm{1}\{ \mathcal{A}(Z_i') > h \}$ \;
Calculate discretized predictions on test data: $\widehat{Y}_j = \mathbbm{1}\{ \mathcal{A}(Z_j) > h \}$ \;
}
\For{$c \in \mathcal{C}$}{
$\bm{\pi}_{train} = \left[ \frac{1}{m} \sum \limits_{i=1}^m (1 - Y_i'), \ \frac{1}{m} \sum \limits_{i=1}^m Y_i' \right]^T $ \;
$\mathbf{M} \in \mathbb{R}^{2 \times 2}$, with $\mathbf{M}_{ij} = \frac{1}{m} \sum \limits_{k=1}^m \mathbbm{1}\{ \widehat{Y}_k' = i-1, Y_k' = j-1 \}$, for $i,j \in \{1, 2\}$ \;
$\widehat{\bm{\pi}}_{test} = \left[\frac{1}{\# \{i: C_i = c\}} \sum \limits_{j=1}^n (1 - \widehat{Y}_j) \mathbbm{1}\{C_j = c\}, \ \frac{1}{\# \{i: C_i = c\}} \sum \limits_{j=1}^n \widehat{Y}_j \mathbbm{1}\{C_j = c\} \right]^T $ \;
$\widehat{P}(Y = 1 | C = c) = (\mathbf{M}^{-1} \widehat{\bm{\pi}}_{test})[2] \cdot \bm{\pi}_{train}[2]$, where $\bm{v}[2]$ denotes the second element of vector $\bm{v}$ \;
}
 \Return{$\widehat{P}(Y = 1 | C = c)$ for each $c \in \mathcal{C}$}
 \caption{Discretization method for label shift prevalence estimation \citep{lipton2018detecting}}
 \label{alg:lipton-ls}
\end{algorithm}

\subsubsection{Fixed point method}

The discretization method converts predicted probabilities in $(0, 1)$ to binary predictions by thresholding. This requires specifying a threshold, and the choice of threshold may impact the resulting prevalence estimates, particularly if the label shift assumption holds only approximately or there are few observations from one class. An alternative is to consider the label shift corrected probabilities $\mathcal{A}_L(z, c)$, calculated from $\mathcal{A}(z)$ via Bayes theorem \eqref{eq:label-shift-correction}. 

Let $\pi_{test, c}$ be a putative value for $P(Y = 1 | C = c)$, and $\mathcal{A}_L^{\pi_{test,c}}(z,c)$ the corrected probabilities using $\pi_{test,c}$ in \eqref{eq:label-shift-correction}. If $\mathcal{A}$ is a calibrated classifier on the training data, and $\pi_{test,c}$ is close to $P(Y = 1 | C = c)$, then under the label shift assumption $\pi_{test,c} \approx \frac{1}{\# \{i: C_i = c\}} \sum \limits_{j=1}^n \mathcal{A}_L^{\pi_{test,c}}(Z_j,C_j) \mathbbm{1}\{C_j = c\}$. The fixed point method considers a range $[a,b] \subset (0, 1)$ of potential values for $P(Y = 1 | C = c)$, and estimates prevalence by
\begin{align}
\widehat{P}(Y = 1 | C = c) = \argmin \limits_{\pi_{test,c} \in [a,b]} \left| \frac{1}{\# \{i: C_i = c\}} \sum \limits_{j=1}^n \mathcal{A}_L^{\pi_{test,c}}(Z_j,C_j) \mathbbm{1}\{C_j = c\} - \pi_{test,c} \right|.
\end{align}
(The restricted range $[a,b]$ is required to avoid trivial solutions). This is essentially the approached proposed by \cite{saerens2002adjusting}, just implemented as a search rather than through iterated EM updates.

\subsection{Posterior sampling for logistic GAMs}

The inference procedure described in Section \ref{sec:inference} requires the ability to sample sample a new classifier function $\mathcal{A}^*$ at each bootstrap step. As bootstrapping the training data and refitting the classifier at each step is potentially very computationally expensive, it may be necessary to use a classifier for which the variability is understood. In particular, we want to resample the full classification function at every step, not just characterize variability at a single point (this requirement is sufficient to construct global confidence bands for the probability function, not just pointwise confidence intervals). As an example, we show here how standard results for logistic GAMs can be incorporated into the bootstrap procedure described above.

Letting $Z_1,...,Z_d$ denote the components of $Z$, the logistic GAM models $P(Y=1|Z)$ by
\begin{align}
\text{logit} \ P(Y=1 | Z) = f_1(Z_1) + \cdots + f_d(Z_d),
\end{align}
where $f_1,...,f_d$ are smooth functions \citep{hastie1990generalized, wood2017generalized}. The model is estimated with a spline fit. Let $\overline{\mathbf{Z}}$ be the design matrix for the spline fit (capturing the intercept and $Z$), let $\boldsymbol{\beta}$ be the spline coefficients, and $\lambda$ the smoothing parameter. The spline fit is penalized by $\lambda \boldsymbol{\beta}^T \mathbf{S} \boldsymbol{\beta}$ where $\lambda$ is the smoothing parameter and $\mathbf{S}$ is the spline smoothing matrix; for example, for smoothing splines $\lambda \boldsymbol{\beta}^T \mathbf{S} \boldsymbol{\beta}$ corresponds to a penalty on the integrated squared second derivative of the smooth. It turns out that this penalty term is equivalent to using the prior distribution $\boldsymbol{\beta} | \lambda \sim N(\mathbf{0}, (\lambda \mathbf{S})^{-})$, where $(\lambda \mathbf{S})^-$ denotes the pseudo-inverse \citep{wood2017generalized}. The posterior distribution of $\boldsymbol{\beta}$, given the training data $(X_i', Z_i', Y_i')$, is approximately
\begin{align}
\label{eq:beta-posterior}
\boldsymbol{\beta} | \{(Z_i', Y_i') \}_{i=1}^m, \lambda \sim N(\widehat{\boldsymbol{\beta}}, (\overline{\mathbf{Z}}^T \mathbf{W} \overline{\mathbf{Z}} + \lambda \mathbf{S})^{-1}),
\end{align}
where $\widehat{\boldsymbol{\beta}}$ are the estimated coefficients and $\mathbf{W}$ is the weights matrix from the last step of penalized iterative reweighted least squares estimation \citep{wood2017generalized}. 

If we use a logistic GAM as our classifier, we can draw a new classifier function $\mathcal{A}^*$ in the bootstrap procedure discussed above by sampling from the posterior distribution of $\boldsymbol{\beta}$. While we expect draws from a posterior to give Bayesian credible intervals, posteriors for spline fits often have good frequentist properties \citep{krivobokova2010simultaneous, wood2017generalized}, and so it is reasonable to sample from the posterior in \eqref{eq:beta-posterior} to construct our frequentist bootstrap confidence intervals.

\subsection{Bootstrap algorithms for inference}

In Section \ref{sec:inference}, we describe semiparametric bootstrap procedures for inference with prevalence $P(Y = y | C = c)$ and class-conditional feature means $\mathbb{E}[X|Y=y, C=c]$. Here we include the detailed algorithms for implementing each bootstrap procedure. Algorithm \ref{alg:bootstrap-ci-prevalence} describes the procedure from Section \ref{sec:prevalence-inference} for constructing a confidence interval for prevalence; Algorithm \ref{alg:bootstrap-ci} constructs confidence intervals for $\mathbb{E}[X|Y=y, C=c]$ using probability-weighted mixed effects models, as in Section \ref{sec:feature-inference}; and Algorithm \ref{alg:mixture-ci} constructs confidence intervals for $\mathbb{E}[X|Y=y, C=c]$ using parametric mixture models assisted by label shift estimation, as in Section \ref{sec:mixture-models}.

\begin{algorithm}
\small
\KwData{Labeled training data $\{(Z_i', Y_i', C_i', K_i')\}_{i=1}^m$ \newline
Unlabeled test data $\{(Z_j, C_j, K_j)\}_{j=1}^n$}
 \SetKwInOut{Input}{Input}
    \SetKwInOut{Output}{Output}

    \Input{Number $B$ of bootstrap samples \newline
    Level $1 - \alpha$ for confidence interval}
    \Output{A confidence interval for $P(Y=1|C = c)$}
    \Init{}{
Train classifier $\mathcal{A}$ with training data $\{(Z_i', Y_i')\}_{i=1}^m$\;
Calculate estimate $\widehat{P}(Y = 1 | C = c)$ with label shift methods (\ref{sec:label-shift-est}) \;
Calculate label shift-corrected predictions $\mathcal{A}_L(Z_i, C_i)$ as in \eqref{eq:label-shift-correction}\;
}
\For{$s= 1,...,B$}{
Sample $(Z_1'^*, Y_1'^*),...,(Z_m'^*, Y_m'^*)$ by resampling rows $(Z_i', Y_i')$ with replacement\;
   Train classifier $\mathcal{A}^*$ on bootstrap sample $(Z_1'^*, Y_1'^*),...,(Z_m'^*, Y_m'^*)$\;
   Sample $(Z_i^*, C_i^*)$ by resampling rows $(Z_i, C_i)$ with replacement\;
   Using $(Z_i'^*, Y_i'^*)$, $\mathcal{A}^*$, and $Z_i^*$, calculate $\widehat{P}_s(Y^* = 1 | C^* = c)$ with label shift methods (\ref{sec:label-shift-est})\;
}
 \Return{$1 - \alpha$ \emph{confidence interval from} $\widehat{P}(Y = 1 | C = c)$ \emph{and the} $\widehat{P}_s(Y^* = 1 | C^* = c)$ \emph{(e.g., a bootstrap percentile interval)}}
 \caption{Semiparametric bootstrap confidence intervals for prevalence, under label shift}
 \label{alg:bootstrap-ci-prevalence}
\end{algorithm}

\begin{algorithm}
\small
\KwData{Labeled training data $\{(X_i', Z_i', Y_i', C_i', K_i')\}_{i=1}^m$ \newline
Unlabeled test data $\{(X_j, Z_j, c_j, K_j)\}_{j=1}^n$}
 \SetKwInOut{Input}{Input}
    \SetKwInOut{Output}{Output}

    \Input{Number $B$ of bootstrap samples \newline
    Level $1 - \alpha$ for confidence interval}
    \Output{A confidence interval for the parameters $\beta_{c,1}$ of the mixed effects model \eqref{eq:weighted-me-model}}
    \Init{}{
Train classifier $\mathcal{A}$ with training data $\{(Z_i', Y_i')\}_{i=1}^m$\;
Calculate label shift-corrected predictions $\mathcal{A}_L(Z_i, C_i)$ as in \eqref{eq:label-shift-correction}\;
Fit the mixed effects model \eqref{eq:weighted-me-model} for $y=0$ and $y=1$, using weights $w_{i,1} = \mathcal{A}_L(Z_i, c_i)$ and $w_{i,0} = 1 - w_{i,1}$. This gives parameter estimates $\widehat{\beta}_{c,y}$ and $\widehat{\omega}^2$, and observed random effects $\widehat{b}_{k}$\;
Define residuals $e_{i} = X_{i} - \widehat{b}_{k}$\;
}
\For{$s= 1,...,B$}{
Sample $(Z_1'^*, Y_1'^*),...,(Z_m'^*, Y_m'^*)$ by resampling rows $(Z_i', Y_i')$ with replacement\;
   Train classifier $\mathcal{A}^*$ on bootstrap sample $(Z_1'^*, Y_1'^*),...,(Z_m'^*, Y_m'^*)$\;
   Sample $(Z_i^*, C_i^*, K_i^*, \mathcal{A}_L(Z_i^*, c_i^*), e_{i}^*)$ by resampling rows $(Z_i, C_i, K_i, \mathcal{A}_L(Z_i, C_i), e_{i})$ with replacement\;
   Sample $Y_i^* \overset{iid}{\sim} \text{Bernoulli}(\mathcal{A}_L(Z_i^*, C_i^*))$ for $i=1,...,n$\;
   Sample $b_{k}^* \overset{iid}{\sim} N(0, \widehat{\omega}^2)$ for $k \in \mathcal{K}$\;
   Generate $X_i^*$ by $X_i^* = e_{i}^* + b_{k}^*$ for $i=1,...,n$\;
   Using $(Z_i'^*, Y_i'^*)$, $\mathcal{A}^*$, and $Z_i^*$, calculate $\widehat{P}(Y_i^* = 1 | c_i^*)$ under the label shift assumption (\ref{sec:label-shift-est})\; 
   $\mathcal{A}_L^*(Z_i^*, C_i^*) = \dfrac{\frac{\widehat{P}(Y_i^* = 1 | C_i^*)}{\widehat{P}(Y_i'^* = 1)} \mathcal{A}^*(Z_i^*)}{\frac{\widehat{P}(Y_i^* = 1 | C_i^*)}{\widehat{P}(Y_i'^* = 1)} \mathcal{A}^*(Z_i^*) + \frac{1- \widehat{P}(Y_i^* = 1 | C_i^*)}{1 - \widehat{P}(Y_i'^* = 1)} (1 - \mathcal{A}^*(Z_i^*))}$\;
   Fit the mixed effects model \eqref{eq:weighted-me-model} with observed data $(X_i^*, C_i^*, K_i^*)$ and weights $w_{i,1}^* = \mathcal{A}_L^*(Z_i^*, C_i^*)$, giving estimates $\widehat{\beta}_{c,1,s}^*$\;
}
 \Return{$1 - \alpha$ \emph{confidence interval from} $\widehat{\beta}_{c,1}$ \emph{and the} $\widehat{\beta}_{c, 1, s}^*$ \emph{(e.g., a bootstrap percentile interval)}}
 \caption{Semiparametric bootstrap confidence intervals with classifier predictions under label shift}
 \label{alg:bootstrap-ci}
\end{algorithm}

\begin{algorithm}
\small
\KwData{Labeled training data $\{(X_i', Z_i', Y_i', C_i', K_i')\}_{i=1}^m$ \newline
Unlabeled test data $\{(X_j, Z_j, C_j, K_j)\}_{j=1}^n$}
 \SetKwInOut{Input}{Input}
    \SetKwInOut{Output}{Output}

    \Input{Number $B$ of bootstrap samples \newline
    Level $1 - \alpha$ for confidence interval}
    \Output{A confidence interval for the parameters $\beta_{c,1}$ of the mixture model \eqref{eq:me-mixture}}
    \Init{}{
Train classifier $\mathcal{A}$ with training data $\{(Z_i', Y_i')\}_{i=1}^m$\;
Using $(Z_i', Y_i')$, $\mathcal{A}$, and $Z_i$, calculate $\widehat{P}(Y_i = 1 | C_i)$ under the label shift assumption (\ref{sec:label-shift-est})\;
Calculate label shift-corrected predictions $\mathcal{A}_L(Z_i, C_i)$ as in \eqref{eq:label-shift-correction}\;
Using $\widehat{P}(Y_i = 1 | C_i)$ as mixing proportions, fit the mixed effects mixture model \eqref{eq:me-mixture}, giving parameter estimates $\widehat{\beta}_{c,y}$ and $\widehat{\omega}_y^2$ and observed random effects $\widehat{b}_{k,0}$, $y \in \{0,1\}$\;
Define residuals $e_{i,0} = X_{i} - \widehat{b}_{k,1}$ and $e_{i,1} = X_{i} - \widehat{b}_{k,0}$\;
}
\For{$s= 1,...,B$}{
Sample $(Z_1'^*, Y_1'^*),...,(Z_m'^*, Y_m'^*)$ by resampling rows $(Z_i', Y_i')$ with replacement\;
   Train classifier $\mathcal{A}^*$ on bootstrap sample $(Z_1'^*, Y_1'^*),...,(Z_m'^*, Y_m'^*)$\;
   Sample $(Z_i^*, C_i^*, K_i^*, \mathcal{A}_L(Z_i^*, C_i^*), e_{i,0}^*, e_{i,1}^*)$ by resampling rows $(Z_i, C_i, K_i, \mathcal{A}_L(Z_i, C_i), e_{i,0}, e_{i,1})$ with replacement\;
   Sample $Y_i^* \overset{iid}{\sim} \text{Bernoulli}(\mathcal{A}_L(Z_i^*, c_i^*))$ for $i=1,...,n$\;
   Sample $b_{k,0}^* \overset{iid}{\sim} N(0, \widehat{\omega}_0^2)$ and $b_{k,1}^* \overset{iid}{\sim} N(0, \widehat{\omega}_1^2)$ for $k \in \mathcal{K}$\;
   Generate $X_i^*$ by $X_i^* = (e_{i,1}^* + b_{k,1}^*)Y_i^* + (e_{i,0}^* + b_{k,0}^*)(1 - Y_i^*)$ for $i=1,...,n$\;
   Using $(Z_i'^*, Y_i'^*)$, $\mathcal{A}^*$, and $Z_i^*$, calculate $\widehat{P}(Y_i^* = 1 | C_i^*)$ under the label shift assumption (\ref{sec:label-shift-est})\; 
  Using $\widehat{P}(Y_i^* = 1 | C_i^*)$ as mixing proportions, fit the mixed effects mixture model \eqref{eq:me-mixture}, giving parameter estimates $\widehat{\beta}_{c,1,s}^*$\;
}
 \Return{$1 - \alpha$ \emph{confidence interval from} $\widehat{\beta}_{c,1}$ \emph{and the} $\widehat{\beta}_{c, 1, s}^*$ \emph{(e.g., a bootstrap percentile interval)}}
 \caption{Semiparametric bootstrap confidence intervals with mixture models under label shift}
 \label{alg:mixture-ci}
\end{algorithm}

\subsection{Mixed effects models with label-dependent random effects}

As we discuss in Section \ref{sec:inference} and Section \ref{sec:simulations}, confidence intervals from the weighted mixed effects model \eqref{eq:weighted-me-model} may lose coverage when random effects are label dependent. In Table \ref{tab:independent-re-results}, we saw this decrease in coverage, and also that an additional variance calibration step can address the issue. Algorithm \ref{alg:bootstrap-ci-independent-re} details the full mixed effects procedure when random effects are label-dependent, including the variance calibration step. Note that variance calibration approximately doubles the time needed to construct a bootstrap confidence interval.

\begin{algorithm}
\small
\KwData{Labeled training data $\{(X_i', Z_i', Y_i', C_i', K_i')\}_{i=1}^m$ \newline
Unlabeled test data $\{(X_j, Z_j, c_j, K_j)\}_{j=1}^n$}
 \SetKwInOut{Input}{Input}
    \SetKwInOut{Output}{Output}

    \Input{Number $B$ of bootstrap samples \newline
    Level $1 - \alpha$ for confidence interval}
    \Output{A confidence interval for the parameters $\beta_{c,1}$ of the mixed effects model \eqref{eq:weighted-me-model}}
    \Init{}{
Train classifier $\mathcal{A}$ with training data $\{(Z_i', Y_i')\}_{i=1}^m$\;
Calculate label shift-corrected predictions $\mathcal{A}_L(Z_i, C_i)$ as in \eqref{eq:label-shift-correction}\;
Fit the mixed effects model \eqref{eq:weighted-me-model} with $y=1$ and weights $w_{i,1} = \mathcal{A}_L(Z_i, c_i)$, giving parameter estimates $\widehat{\beta}_{c,1}$ and $\widehat{\omega}_1^2$, and observed random effects $\widehat{b}_{k,1}$\;
Fit the mixed effects model \eqref{eq:weighted-me-model} with $y=0$ and weights $w_{i,0} = 1 - w_{i,1}$, giving parameter estimates $\widehat{\beta}_{c,0}$ and $\widehat{\omega}_0^2$, and observed random effects $\widehat{b}_{k,0}$\;
Define residuals $e_{i,0} = X_{i} - \widehat{b}_{k,1}$ and $e_{i,1} = X_{i} - \widehat{b}_{k,0}$\;
}
\For{$s= 1,...,B$}{
Sample $(Z_1'^*, Y_1'^*),...,(Z_m'^*, Y_m'^*)$ by resampling rows $(Z_i', Y_i')$ with replacement\;
   Train classifier $\mathcal{A}^*$ on bootstrap sample $(Z_1'^*, Y_1'^*),...,(Z_m'^*, Y_m'^*)$\;
   Sample $(Z_i^*, C_i^*, K_i^*, \mathcal{A}_L(Z_i^*, c_i^*), e_{i,0}^*, e_{i,1}^*)$ by resampling rows $(Z_i, C_i, K_i, \mathcal{A}_L(Z_i, C_i), e_{i,0}, e_{i,1})$ with replacement\;
   Sample $Y_i^* \overset{iid}{\sim} \text{Bernoulli}(\mathcal{A}_L(Z_i^*, C_i^*))$ for $i=1,...,n$\;
   Sample $b_{k,0}^* \overset{iid}{\sim} N(0, \widehat{\omega}_0^2)$ and $b_{k,1}^* \overset{iid}{\sim} N(0, \widehat{\omega}_1^2)$ for $k \in \mathcal{K}$\;
   Generate $X_i^*$ by $X_i^* = (e_{i,1}^* + b_{k,1}^*)Y_i^* + (e_{i,0}^* + b_{k,0}^*)(1 - Y_i^*)$ for $i=1,...,n$\;
   Using $(Z_i'^*, Y_i'^*)$, $\mathcal{A}^*$, and $Z_i^*$, calculate $\widehat{P}(Y_i^* = 1 | c_i^*)$ under the label shift assumption (\ref{sec:label-shift-est})\; 
   $\mathcal{A}_L^*(Z_i^*, C_i^*) = \dfrac{\frac{\widehat{P}(Y_i^* = 1 | C_i^*)}{\widehat{P}(Y_i'^* = 1)} \mathcal{A}^*(Z_i^*)}{\frac{\widehat{P}(Y_i^* = 1 | C_i^*)}{\widehat{P}(Y_i'^* = 1)} \mathcal{A}^*(Z_i^*) + \frac{1- \widehat{P}(Y_i^* = 1 | C_i^*)}{1 - \widehat{P}(Y_i'^* = 1)} (1 - \mathcal{A}^*(Z_i^*))}$\;
   Fit the mixed effects model \eqref{eq:weighted-me-model} with observed data $(X_i^*, C_i^*, K_i^*)$ and weights $w_{i,1}^* = \mathcal{A}_L^*(Z_i^*, C_i^*)$ and $w_{i,0}^* = 1 - w_{i,1}^*$, giving estimates $\widehat{\omega}_{1,s}^{*2}$ and $\widehat{\omega}_{0,s}^{*2}$\;
   Calculate the true sample variance $v_{y,s}^2 = \frac{1}{|\mathcal{K}| - 1} \sum \limits_k b_{k,y}^{*2}$ \;
}
Regress $v_{y,s}^2$ on $\widehat{\omega}_{y,s}^{*2}$, producing an estimating function $\widehat{f}_y$ with $\widehat{v}_{y,s}^2 = \widehat{f}_y(\widehat{\omega}_{y,s}^{*2})$, for $y \in \{0,1\}$ \;
Calculate the adjusted variances: $\widehat{\omega}_{y, \text{adj}}^2 = \widehat{f}_y(\widehat{\omega}_y^2)$, for $y \in \{0,1\}$ \;
\For{$s= 1,...,B$}{
Sample $(Z_1'^*, Y_1'^*),...,(Z_m'^*, Y_m'^*)$ by resampling rows $(Z_i', Y_i')$ with replacement\;
   Train classifier $\mathcal{A}^*$ on bootstrap sample $(Z_1'^*, Y_1'^*),...,(Z_m'^*, Y_m'^*)$\;
   Sample $(Z_i^*, C_i^*, K_i^*, \mathcal{A}_L(Z_i^*, c_i^*), e_{i,0}^*, e_{i,1}^*)$ by resampling rows $(Z_i, C_i, K_i, \mathcal{A}_L(Z_i, C_i), e_{i,0}, e_{i,1})$ with replacement\;
   Sample $Y_i^* \overset{iid}{\sim} \text{Bernoulli}(\mathcal{A}_L(Z_i^*, C_i^*))$ for $i=1,...,n$\;
   Sample $b_{k,0}^* \overset{iid}{\sim} N(0, \widehat{\omega}_{0, \text{adj}}^2)$ and $b_{k,1}^* \overset{iid}{\sim} N(0, \widehat{\omega}_{1, \text{adj}}^2)$ for $k \in \mathcal{K}$\;
   Generate $X_i^*$ by $X_i^* = (e_{i,1}^* + b_{k,1}^*)Y_i^* + (e_{i,0}^* + b_{k,0}^*)(1 - Y_i^*)$ for $i=1,...,n$\;
   Using $(Z_i'^*, Y_i'^*)$, $\mathcal{A}^*$, and $Z_i^*$, calculate $\widehat{P}(Y_i^* = 1 | c_i^*)$ under the label shift assumption (\ref{sec:label-shift-est}) \; 
   $\mathcal{A}_L^*(Z_i^*, C_i^*) = \dfrac{\frac{\widehat{P}(Y_i^* = 1 | C_i^*)}{\widehat{P}(Y_i'^* = 1)} \mathcal{A}^*(Z_i^*)}{\frac{\widehat{P}(Y_i^* = 1 | C_i^*)}{\widehat{P}(Y_i'^* = 1)} \mathcal{A}^*(Z_i^*) + \frac{1- \widehat{P}(Y_i^* = 1 | C_i^*)}{1 - \widehat{P}(Y_i'^* = 1)} (1 - \mathcal{A}^*(Z_i^*))}$\;
   Fit the mixed effects model \eqref{eq:weighted-me-model} with observed data $(X_i^*, C_i^*, K_i^*)$ and weights $w_{i,1}^* = \mathcal{A}_L^*(Z_i^*, C_i^*)$, giving estimates $\widehat{\beta}_{c,1,s}^*$\;
}
 \Return{$1 - \alpha$ \emph{confidence interval from} $\widehat{\beta}_{c,1}$ \emph{and the} $\widehat{\beta}_{c, 1, s}^*$ \emph{(e.g., a bootstrap percentile interval)}}
 \caption{Semiparametric bootstrap confidence intervals with classifier predictions under label shift, with variance calibration step}
 \label{alg:bootstrap-ci-independent-re}
\end{algorithm}

\subsection{Simulation settings}

In Section \ref{sec:simulations}, we conduct simulations to assess the impact of assumptions on the coverage of our bootstrap confidence intervals. We describe three different scenarios, with different combinations of (A1), (A2), and (A3) holding true. We also assess the impact of deviations from parametric assumptions on the mixture model approach. In Table \ref{tab:sim-settings} we provide the simulation settings, describing how training and test data was simulated in each scenario. Note that $N(\mu, \sigma^2)$ denotes a Gaussian distribution with mean $\mu$ and variance $\sigma^2$, and $SN(\xi, \omega, \alpha)$ denotes a skewed normal distribution with location $\xi$, scale $\omega$, and shape $\alpha$.

\begin{table}[]
\centering
\begin{tabular}{|c|c|c|c|}
\hline
Assumptions                       & Normal? & Training Data                                                                                                                                         & Test Data                                                                                                                                                                                                       \\ \hline
\multirow{2}{*}{(A1), (A2), (A3)} & yes     & \begin{tabular}[c]{@{}c@{}}$Y' \sim Bernoulli(0.2)$\\ $Z' | Y' = 0 \sim N(0, 1)$\\ $Z' | Y' = 1 \sim N(3, 1)$\end{tabular}                            & \begin{tabular}[c]{@{}c@{}}$Y \sim Bernoulli(0.4)$ \\ $Z | Y = 0 \sim N(0, 1)$\\ $Z | Y = 1 \sim N(3, 1)$\\ $b_k \sim N(0, 0.5)$\\ $X_i = Z_i + b_{k_i} + N(0, 0.2)$\end{tabular}                               \\ \cline{2-4} 
                                  & no      & \begin{tabular}[c]{@{}c@{}}$Y' \sim Bernoulli(0.2)$\\ $Z' | Y' = 0 \sim SN(0, 2, 3)$\\ $Z' | Y' = 1 \sim 8 - SN(3, 2, 3)$\end{tabular}                & \begin{tabular}[c]{@{}c@{}}$Y \sim Bernoulli(0.4)$\\ $Z | Y = 0 \sim SN(0, 2, 3)$\\ $Z | Y = 1 \sim 8 - SN(3, 2, 3)$\\ $b_k \sim N(0, 0.5)$\\ $X_i = Z_i + b_{k_i} + N(0, 0.2)$\end{tabular}                    \\ \hline
\multirow{2}{*}{(A2), (A3)}       & yes     & \begin{tabular}[c]{@{}c@{}}$Y' \sim Bernoulli(0.2)$\\ $Z' | Y' = 0 \sim N(-0.5, 1)$\\ $Z' | Y' = 1 \sim N(3, 1)$\end{tabular}                         & \begin{tabular}[c]{@{}c@{}}$Y \sim Bernoulli(0.4)$ \\ $Z | Y = 0 \sim N(0, 1)$\\ $Z | Y = 1 \sim N(3, 1)$\\ $b_k \sim N(0, 0.5)$\\ $X_i = Z_i + b_{k_i} + N(0, 0.2)$\end{tabular}                               \\ \cline{2-4} 
                                  & no      & \begin{tabular}[c]{@{}c@{}}$Y' \sim Bernoulli(0.2)$\\ $Z' | Y' = 0 \sim SN(-0.5, 2, 3)$\\ $Z' | Y' = 1 \sim 8 - SN(3, 2, 3)$\end{tabular}             & \begin{tabular}[c]{@{}c@{}}$Y \sim Bernoulli(0.4)$\\ $Z | Y = 0 \sim SN(0, 2, 3)$\\ $Z | Y = 1 \sim 8 - SN(3, 2, 3)$\\ $b_k \sim N(0, 0.5)$\\ $X_i = Z_i + b_{k_i} + N(0, 0.2)$\end{tabular}                    \\ \hline
\multirow{2}{*}{(A1), (A2)}       & yes     & \begin{tabular}[c]{@{}c@{}}$Y' \sim Bernoulli(0.2)$\\ $Z' | Y' = 0 \sim N(0, 1)$\\ $Z' | Y' = 1 \sim N(3, 1)$\end{tabular}                            & \begin{tabular}[c]{@{}c@{}}$Y \sim Bernoulli(0.4)$ \\ $Z | Y = 0 \sim N(0, 1)$\\ $Z | Y = 1 \sim N(3, 1)$\\ $b_k \sim N(0, 0.5)$\\ $X_i = Z_i + b_{k_i} + N(0, 0.2) + \mathbbm{1}(Y_i = 1)$\end{tabular}            \\ \cline{2-4} 
                                  & no      & \begin{tabular}[c]{@{}c@{}}$Y' \sim Bernoulli(0.2)$\\ $Z' | Y' = 0 \sim SN(0, 2, 3)$\\ $Z' | Y' = 1 \sim 8 - SN(3, 2, 3)$\end{tabular} & \begin{tabular}[c]{@{}c@{}}$Y \sim Bernoulli(0.4)$\\ $Z | Y = 0 \sim SN(0, 2, 3)$\\ $Z | Y = 1 \sim 8 - SN(3, 2, 3)$\\ $b_k \sim N(0, 0.5)$\\ $X_i = Z_i + b_{k_i} + N(0, 0.2) + \mathbbm{1}(Y_i = 1)$\end{tabular} \\ \hline
\end{tabular}
\caption{Simulation settings for assessing performance of bootstrap inference procedures.}
\label{tab:sim-settings}
\end{table}

\subsection{Functional PCA features}

After automatic event detection and particle tracking, each detected event on the cell surface is represented by a series of greyscale $9 \times 9$ pixel frames, with the intensity of fluorescence recorded for each pixel. As shown in Figure \ref{fig:tirf}, there are often clear differences in the individual frames for puffs and nonpuffs, and more importantly in the evolution of frames over time.

\begin{figure}
\centering
\includegraphics[scale=0.35]{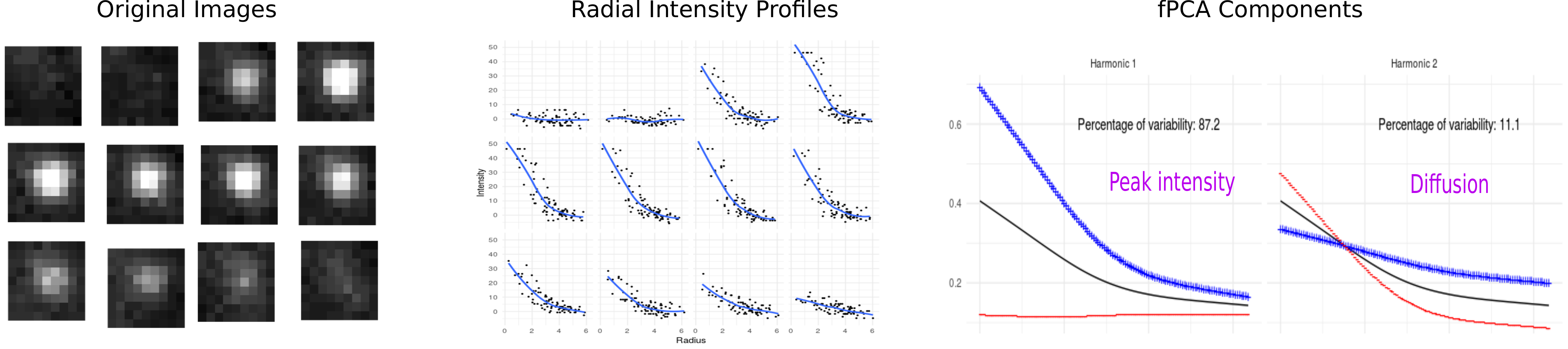}
\caption{Functional PCA on radial intensity profiles. \underline{Left}: the original frames for a puff from TIRF microscopy. \underline{Center}: the corresponding radial intensity profiles for the frames on the left, with smooth curves showing the overall shape in each frame. \underline{Right}: the first two fPCA component functions for the radial intensity profiles, plotted as differences from the mean function (the black curve). A positive score for the component function is plotted with blue $+$'s, while a negative score for the component function is plotted with red $-$'s.}
\label{fig:fpca}
\end{figure}

To capture the behavior of puffs over time, we first consider the two-dimensional intensity function within each frame. Noticing that puffs tend to have symmetric intensity functions, we can reduce the intensity function to a one-dimensional function of distance from the center of the frame (Figure \ref{fig:fpca}). Each event then becomes a collection of radial intensity functions. As demonstrated in Figure \ref{fig:fpca}, the radial intensity functions tend to have similar shapes, with only a few modes of variation. This suggests that functional PCA (fPCA) \citep{ramsay2005fda} could provide effective dimension reduction of the radial profiles.

\begin{figure}
\centering
\includegraphics[scale=0.4]{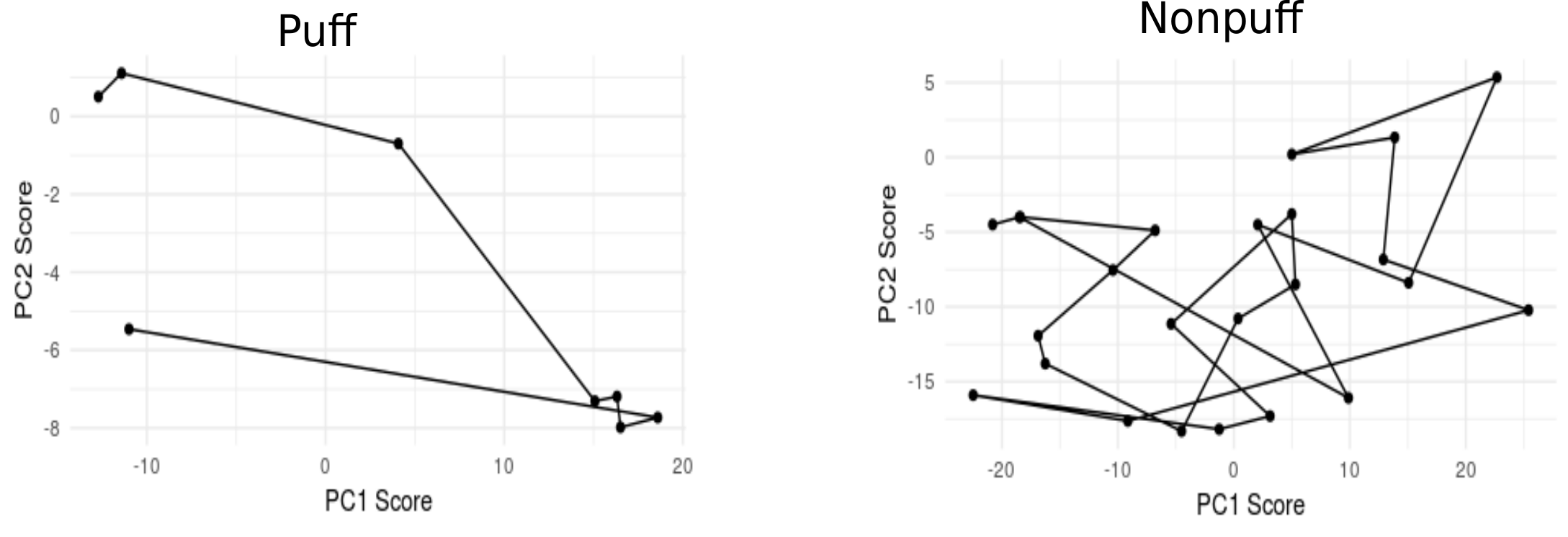}
\caption{Differences between fPCA score paths for puffs and nonpuffs.}
\label{fig:fpca-score-path}
\end{figure}

Before performing fPCA, each detected event was scaled to have the same peak intensity. Figure \ref{fig:fpca} shows the first two principal component functions, which together make up about 98\% of the variability in radial intensity profiles. As we might expect, peak intensity in a frame is the main component of variation, captured by the first principal component. The second principal component, account for about 11\% of the variability in radial intensity profiles, captures diffusivity of fluorescence in the frame. Each event is then represented as a bivariate time series of scores for the first two principal components. As suggested by Figure \ref{fig:tirf}, time series of component scores for puffs are expected to have a characteristic pattern, whereas scores for nonpuffs are expected to be much noisier. To visualize fPCA score time series, we consider each event as a path through two-dimensional principal component score space. Figure \ref{fig:fpca-score-path} shows these paths for a puff and a nonpuff, illustrating the differences we expect to see in these time series.

Figure \ref{fig:fpca-score-path} suggests that featurizing the fPCA score paths could be useful for classifying puffs and nonpuffs. We construct several features:
\begin{itemize}
\item $ConvexArea$ and $ConvexPerimeter$: the area and perimeter of a convex hull around the score path

\item $Noise$: a measure of randomness in the time series of first principal component scores

\item $Smoothness$: the average distance between points in the observed score path and the kernel-smoothed score path
\end{itemize}

\subsection{TIRF microscopy mixture model results}

\begin{figure}
\centering
\includegraphics[scale=0.45]{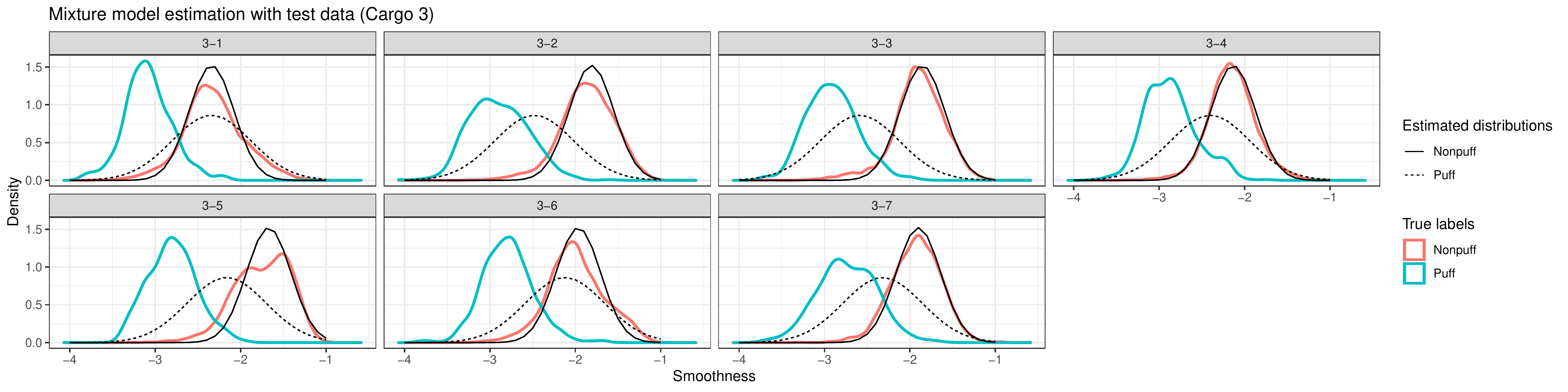}
\caption{The distribution of $Smoothness$ in each cell, for puffs and nonpuffs. Estimates of the true densities are shown from kernel density estimation with the true labels, while the black curves show the fitted normal distributions from a hierarchical Gaussian mixture model. For the mixture model, the mixing proportion for puffs and nonpuffs is estimated as a parameter of the model.}
\label{fig:bad-mixture}
\end{figure}

\begin{figure}
\centering
\includegraphics[scale=0.45]{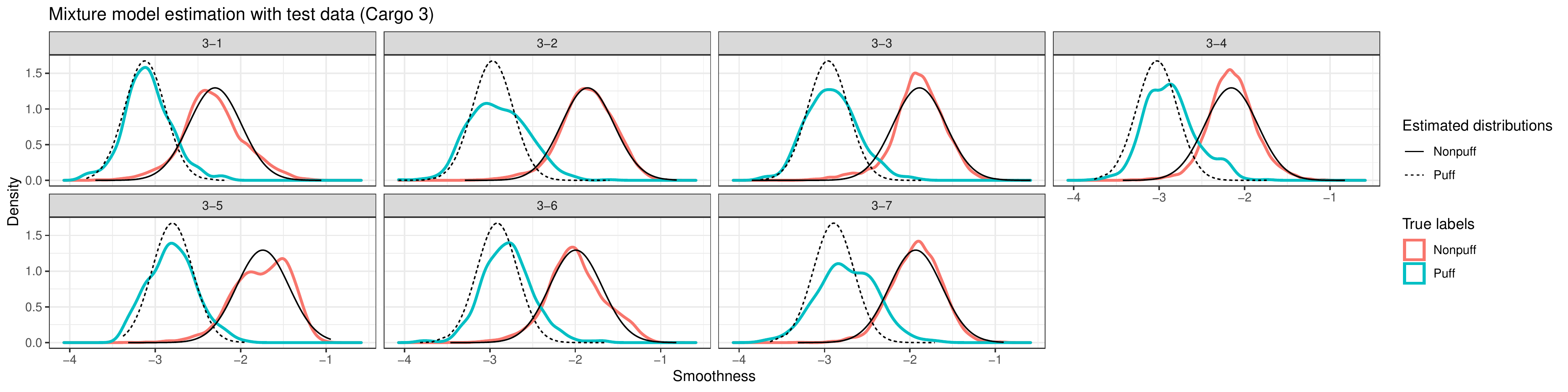}
\caption{The distribution of $Smoothness$ in each cell, for puffs and nonpuffs. Estimates of the true densities are shown from kernel density estimation with the true labels, while the black curves show the fitted normal distributions from a hierarchical Gaussian mixture model. For the mixture model, the mixing proportion in each cell is estimated beforehand with the label shift correction, to improve identifiability.}
\label{fig:good-mixture}
\end{figure}

In Section \ref{sec:case-study}, we summarize the results of inference on puff $Smoothness$, using the mixed effects approach and the mixture model approach. The full results are provided here. Figure \ref{fig:bad-mixture} shows the distribution of $Smoothness$ within each test cell, for puffs and nonpuffs. The distributions in Figure \ref{fig:bad-mixture} suggest that a two-component Gaussian mixture is reasonable, and so we fit the hierarchical mixture described in \eqref{eq:me-mixture}. However, because the parametric model does not hold exactly, and the proportion of puffs is small, the estimated puff distributions are poor (Figure \ref{fig:bad-mixture}). With the mixing proportions specified, Figure
\ref{fig:good-mixture} shows the resulting fit, which is much improved over
Figure \ref{fig:bad-mixture}. The estimated means in each cell are given by
Table \ref{tab:mixture-model-estimates}; the slight bias in puff means arises
because the distribution of $Smoothness$ is only approximately Gaussian.

\begin{table}[]
\centering
\begin{tabular}{c|cccc}
\hline
\multirow{2}{*}{Cell} & \multicolumn{2}{c}{True means} & \multicolumn{2}{c}{Mixture model estimates} \\
                      & Puff          & Nonpuff        & Puff                & Nonpuff               \\
                      \hline
3-1              & -3.08         & -2.32          & -3.12               & -2.30                 \\
3-2              & -2.90         & -1.86          & -2.97               & -1.86                 \\
3-3              & -2.91         & -1.90          & -2.96               & -1.89                 \\
3-4              & -2.87         & -2.15          & -3.03               & -2.16                 \\
3-5              & -2.80         & -1.75          & -2.80               & -1.74                 \\
3-6              & -2.79         & -1.99          & -2.91               & -1.99                 \\
3-7              & -2.73         & -1.93          & -2.89               & -1.93    \\           
\hline 
\end{tabular}
\caption{Mean of $Smoothness$ in each Condition 3 cell. The true means are estimated with the true labels, while the mixture model estimates result from a mixed effects Gaussian mixture model \eqref{eq:me-mixture}, where the mixing proportion in each cell is estimated with the label shift correction. There is slight bias in the mixture model estimates of puff means shown here; this occurs because the distribution of $Smoothness$ for puffs is only approximately Gaussian, and is slightly right-skewed (Figure \ref{fig:good-mixture}).}
\label{tab:mixture-model-estimates}
\end{table}

\end{document}